\documentclass[prx,twocolumn,english,amsmath,amssymb,superscriptaddress,floatfix,longbibliography]{revtex4-1} 
\usepackage{float}
\usepackage{graphicx}
\usepackage{dcolumn}
\usepackage{bm}
\usepackage{xcolor}

\newcommand*{\cB}{\mathcal{B}}

\newcommand*{\cD}{\mathcal{D}}
\newcommand*{\cE}{\mathcal{E}}
\newcommand*{\cF}{\mathcal{F}}

\newcommand*{\cZ}{\mathcal{Z}}

\newcommand*{\ep}{\epsilon}

\newcommand*{\B}{\bm{B}}

\newcommand*{\BDB}{\B\cdot \dl B}

\renewcommand*{\u}{\bm{u}}

\newcommand*{\Jpl}{J_{\|}}

\newcommand*{\dl}{\bm{\nabla}}
\newcommand*{\del}{\partial}
\newcommand*{\BD}{\bm{B}\cdot\bm{\nabla}}
\newcommand*{\uD}{\bm{u}\cdot\bm{\nabla}}

\newcommand*{\lbr}{\left(}
\newcommand*{\rbr}{\right)}

\newcommand*{\etab}{\overline{\eta}}

\newcommand*\at[2]{\left.#1\right|_{#2}}

\makeatletter
\newsavebox{\@brx}
\newcommand{\llangle}[1][]{\savebox{\@brx}{\(\m@th{#1\langle}\)}%
  \mathopen{\copy\@brx\mkern2mu\kern-0.9\wd\@brx\usebox{\@brx}}}
\newcommand{\rrangle}[1][]{\savebox{\@brx}{\(\m@th{#1\rangle}\)}%
  \mathclose{\copy\@brx\mkern2mu\kern-0.9\wd\@brx\usebox{\@brx}}}
\makeatother

\begin{document}

\preprint{AIP/123-QED}

\title[Sengupta et al.]{Periodic Korteweg-de Vries soliton potentials generate quasisymmetric magnetic fields}

\author{W. Sengupta}
 \altaffiliation[Email: ]{wsengupta@princeton.edu}
 \affiliation{ 
Department of Astrophysical Sciences, Princeton University, Princeton, NJ, 08543
}

\author{N. Nikulsin}
 \affiliation{ 
Department of Astrophysical Sciences, Princeton University, Princeton, NJ, 08543
}
 \affiliation{ 
Department of Astrophysical Sciences, Princeton University, Princeton, NJ, 08543
}
\affiliation{%
Stellarator Theory Department, Max Plank Institute for Plasma Physics, 17491 Greifswald, Germany%
}

\author{S. Buller}
\affiliation{ 
Department of Astrophysical Sciences, Princeton University, Princeton, NJ, 08543
}

\author{R. Madan}
 \affiliation{ 
Department of Astrophysical Sciences, Princeton University, Princeton, NJ, 08543
}
\affiliation{%
Princeton Plasma Physics Laboratory, Princeton, NJ, 08540
}%

\author{E.J. Paul}
 \affiliation{ %
Columbia University, New York, NY 10027, USA%
}

\author{R. Nies}
 \affiliation{ %
Department of Astrophysical Sciences, Princeton University, Princeton, NJ, 08543
}
\affiliation{%
Princeton Plasma Physics Laboratory, Princeton, NJ, 08540
}%

\author{A.A. Kaptanoglu}
\affiliation{Courant Institute of Mathematical Sciences, New York University, New York, NY 10012, USA}

\author{S.R. Hudson}
\affiliation{%
Princeton Plasma Physics Laboratory, Princeton, NJ, 08540
}%

\author{A. Bhattacharjee}
 \affiliation{ %
Department of Astrophysical Sciences, Princeton University, Princeton, NJ, 08543
}

\date{\today}

\begin{abstract}
 
Quasisymmetry (QS) is a hidden symmetry of the magnetic field strength, $B=|{\B}|$, that effectively confines charged particles in a three-dimensional toroidal plasma equilibrium. Here, we show that QS has a deep connection to the underlying symmetry that makes solitons possible. Our approach uncovers a hidden lower dimensionality of $B$ on a magnetic flux surface, which could make stellarator optimization schemes significantly more efficient. Recent numerical breakthroughs (M. Landreman and E. Paul, \textit{Phys. Rev. Lett.} \textbf{128}, 035001 (2022)) have yielded configurations with excellent volumetric QS and surprisingly low magnetic shear. Given $B$, it may be possible to deduce an upper bound on the maximum quasisymmetric toroidal volume which depends only on the properties of $B$. This has been verified for the Landreman-Paul precise quasiaxisymmetric (QA) stellarator configuration. In the neighborhood of the outermost surface, we show that $B$ approaches the form of the 1-soliton reflectionless potential (I. Gjaja and A. Bhattacharjee, \textit{Phys. Rev. Lett.} \textbf{68}, 2413 (1992)). The connection length diverges, indicating the possible presence of an X-point or cusp that could potentially be used as a basis for a divertor. We present a non-perturbative approach based on ensuring single-valuedness of $B$, which directly leads to its Painlev\'e property and the KdV and Gardner's equations. Finally, we use an approach based on machine learning, trained on a large dataset of numerically optimized quasisymmetric stellarators. We robustly recover the KdV and Gardner's equations from the data. 

\end{abstract}

\maketitle

\section{Introduction}
Physical systems possessing a hidden symmetry attract serious attention 
\cite{HS_charalambous2021,HS_lee2021hidden,HS_liu2022machine,landreman2021,moser1979hidden_symmetries,leach2008ermakov_commentary} because of the far-reaching impact of the symmetry on the physical properties of the systems. The hidden symmetry manifests only in special coordinates, which are constructed, along with the invariants, as part of the solution to the problem. In the presence of hidden symmetries, the partial differential equations (PDEs) governing classical dynamical fields often reduce to integrable PDEs 
\citep{moser1979hidden_symmetries,ovsienko_Khesin1987korteweg,leach2008ermakov_commentary}.
An example of the latter is the well-known Korteweg-de Vries (KdV) equation with soliton solutions \cite{GGKM_1967_KdV,novikov1984theory_solitons}.

In plasma physics, a hidden symmetry called \textit{quasisymmetry} (QS)
\citep{boozer1983, Helander2014}, forces the time-independent magnetic field strength, $B=|\B|$, to be symmetric  \citep{boozer1983,rodriguez2022_thesis}, whereas the full vector magnetic field $\B$ itself depends on all three spatial coordinates, and is thus said to be three-dimensional (3D). The similarity of this symmetry to particle relabelling symmetries \citep{padhye1996relabeling,webb2018magnetohydrodynamics_book} is known  \citep{rodriguez2022_thesis,weitzner2021steadyflow} but yet to be fully explored. 

The hidden symmetry in $B$ makes the bounce action (second adiabatic invariant), $\Jpl$, independent of the field-line label \citep{Helander2014},
which guarantees excellent particle confinement in a torus.
The 3D nature of \textbf{B} allows stellarators enough flexibility to avoid intrinsic difficulties (such as powerful transients) associated with large plasma currents in axisymmetric configurations such as tokamaks.

It is currently unknown whether exact QS can be achieved in a toroidal volume for a magnetic field that satisfies the constraint of force balance in ideal magnetohydrodynamics (MHD) \cite{freidberg2014idealMHD}. The problem is of fundamental importance and potentially significant for the design of next-generation stellarator fusion reactors.
Analytical results \citep{garrenboozer1991a,garrenboozer1991b,plunk2018,landreman2019} seem to indicate that in the presence of scalar plasma pressure, the resulting 3D nonlinear system of PDEs does not have solutions in a toroidal volume \citep{garrenboozer1991a,garrenboozer1991b,landreman2019,constantin2021,rodriguez2020a,rodriguez2021weak}. In particular, the near-axis expansion (NAE), which is an asymptotic expansion in the distance from the magnetic axis \citep{mercier1974lectures, Solovev1970,bernardin1986}, when applied to QS \citep{garrenboozer1991a,garrenboozer1991b,landreman2018a}, becomes overdetermined beyond the second order. Moreover, NAEs are typically divergent \citep{frankfu2025NAE} beyond second order. 

Overdetermined problems do not generally have solutions. Even when they have solutions, they may still be subject to restrictive constraints. For example, in the particular case of isodynamic magnetic fields, which satisfy a constraint more restrictive than QS \citep{Helander2014,schief2003nested}, it has been shown \citep{palumbo1968,schief2003nested} that if a solution exists, it must be essentially 2D: either axisymmetric like a tokamak or a helically symmetric straight stellarator.

On the other hand, numerical 3D toroidal stellarator solutions with a precise level of QS and MHD force balance have been obtained via numerical optimization, suggesting the existence of special, physically interesting toroidal solutions to the overdetermined problem. However, the numerical search over a typically vast parameter space obscures insight into the quasisymmetric configuration space that permits precise solutions. In particular, optimized quasisymmetric configurations tend to form clusters in parameter space \citep{rodriguez2022topology,giulianiRodriguez_spivak2024comprehensive}, and may be due to a hidden lower-dimensionality that warrants further analysis.

To construct an analytical model of quasisymmetric vacuum fields in a generic non-symmetric toroidal domain, we must first discuss the fundamental requirements of such a theory. First and foremost, we require global toroidal solutions for $B$ that respect double periodicity (both poloidal and toroidal). We restrict ourselves to the class of smooth magnetic fields that are assumed to be analytic. The analyticity requirement can be rigorously justified near the axis \citep{garrenboozer1991a,landreman2018a, SenguptaNAE_allorders_2024}. Our analyticity requirement currently excludes interesting classes such as piecewise omnigeneous stellarators \citep{velasco2024PRLpiecewise}, which can have sharp changes in the gradients of the field strength.

Second, we require field-line independence of $\Jpl$, as part of the definition of the properties of QS and \text{omnigeneity}. 
The connection with the KdV equation comes from identifying the field-line label $\alpha$ as time. With this identification, we further require that the adiabatic invariant $\Jpl$ be an exact (time-independent) invariant. Most adiabatic invariants with analytic time-dependent potentials are not exact. For example, it can be shown that over a long time, adiabatic invariants change by  \citep{landau1960mechanics,henrard1993adiabatic} equal to $C\exp{(-1/\ep)}$, where $C$ is a positive $\ep$-dependent constant where $\ep\ll 1$ is the slowness or the adiabatic parameter. Fortunately, there exist privileged time-dependent potentials for which the coefficient $C$ multiplying the exponentially small term vanishes identically. Consequently, the adiabatic invariants become exact invariants \citep{gjaja_Bhattacharjee_1992asymptotics_reflectionless,berry_Howles_1990fake_Airy_reflectionless}. Since $C$ is proportional to the reflection coefficient in the case of above-barrier reflection from a potential well \citep{keller1991adiabatic_changes,gjaja_Bhattacharjee_1992asymptotics_reflectionless,berry_Howles_1990fake_Airy_reflectionless}, such potentials are also called reflectionless potentials. Reflectionless potentials are intimately connected with solitons and integrable systems \citep{ablowitz_2011_book_nonlinear_waves,novikov1984theory_solitons} where they naturally appear in the long wavelength limit of periodic cnoidal waves. 

Broadening the search to other integrable periodic nonlinear waves leads us to the question, posed by R.S. MacKay, as to whether Painlev\'e analysis could determine non-axisymmetric QS equilibrium \footnote{R.~MacKay, private communication (Simons team meeting 2019, The Painlev\'e property (PP) should perhaps also be referred to as the Kovalevskaya-Painlev\'e property since Kovalevskaya used it to find new cases of integrable tops before Painlev\'e's work was published (Mich\'ele Audin, \textit{Spinning tops: A course on integrable systems}, Vol. 51 (Cambridge University Press, 1999)).}. The Painlev\'e analysis, which involves the study of singularities of differential equations in the complex domain, has yielded fundamental results in several physics problems from quantum field theory to optics \citep{levi2013painleve}. Analytic continuation from the real to the complex plane is a helpful technique that physicists often utilize in solving linear problems. For nonlinear systems, Kovaleskaya showed that continuation to the complex plane and analyzing singularities can be equally insightful. Moreover, singularities in the complex plane, even those that are far from the real line, can determine and control the behavior of analytic functions on the real line. An important example of this is the doubly periodic elliptic function \citep{hille1997_Complex_ODE,whittaker_watson_1920course} for which, singularities determine its periodicity on the real line.

In our context, the Painlev\'e analysis offers a useful methodology for ensuring global analyticity and single-valuedness of the quasisymmetric solutions in the toroidal domain, which is a highly nontrivial problem. The Painlev\'e property, shared by reflectionless and soliton potentials such as cnoidal waves, guarantees robust exact invariants, global analyticity, and the single-valuedness of $B$ in a toroidal domain. Although it is a priori not obvious that quasisymmetric $B$ belongs to such a restricted class of potentials, the cnoidal solution is entirely consistent with the NAE approach within its limit of applicability.

In this work, we demonstrate that QS admits a class of analytic $B$ that is a periodic solution \citep{novikov1984theory_solitons,kamchatnov2000nonlinear} of the KdV equation. Starting from the basic definitions and properties of QS and requiring only the analyticity and single-valuedness of $B$, we show, non-perturbatively, the connection to the theory of periodic KdV solitons. In making the connection to soliton theory, we demonstrate how the field-line label independence of the second adiabatic invariant $\Jpl$ and other integrals of $B$ in QS are directly related to the infinite hierarchy of conserved quantities of the KdV equation. The exact analytic solution for $B$ is valid in the entire toroidal volume and reduces to the NAE predictions close to the axis. 

Our final approach is data-driven. We use PySINDy \cite{desilva2020, Kaptanoglu2022} to look for relations between $B$ and its derivatives in the large family of quasi-axisymmetric configurations generated in \cite{buller2024family} and the QUASR database \cite{giuliani2023direct}. PySINDy employs sparse regression techniques and algorithms to discover governing dynamical system models from noisy, limited data. We show that, over a wide range of configurations and parameters, the KdV equation with the traveling-wave form for $B$ holds. As described later in the paper, the only exceptions we found are covered by Gardner's equation.

The connection with periodic KdV solitons allows us to describe $B$ on a magnetic flux surface with only three functions of the magnetic flux, which are related to the so-called \textit{spectral parameters} \citep{novikov1984theory_solitons} of the KdV equation. We show that these functions are closely related to the spectrum of $B$ in Boozer coordinates \cite{Helander2014} where the hidden symmetry is manifest, thereby highlighting the low dimensionality of the Boozer spectrum. Furthermore, our analysis shows that the periodicity of $B$ is maintained only when these functions are distinct. Consequently, it may be possible to estimate the region of validity of volumetric QS based on where the two spectral parameters intersect (discussed below). Remarkably, near the point of intersection, the connection length (period of $B$) begins to diverge, and $B$ approaches the form of a reflectionless potential (a single soliton that decays at infinity) \citep{gjaja_Bhattacharjee_1992asymptotics_reflectionless,berry_Howles_1990fake_Airy_reflectionless}. The diverging connection length is a characteristic of an X-point or sharp ridge that could potentially be the basis for a non-resonant divertor \citep{mioduszewski2007power,boozer2015,punjabi2020}.

We demonstrate, using both data-driven and traditional numerical methods, that our analytic theory can describe well the properties of the field strength on a flux surface of the vacuum equilibrium with precise quasi-axisymmetry (QA) \cite{landreman2021} and some other optimized quasisymmetric configurations as well.

\section{Basic formulation of quasisymmetry}
\label{sec:basic_QS}
\begin{figure}
    \centering
    \includegraphics[width=0.45\textwidth]{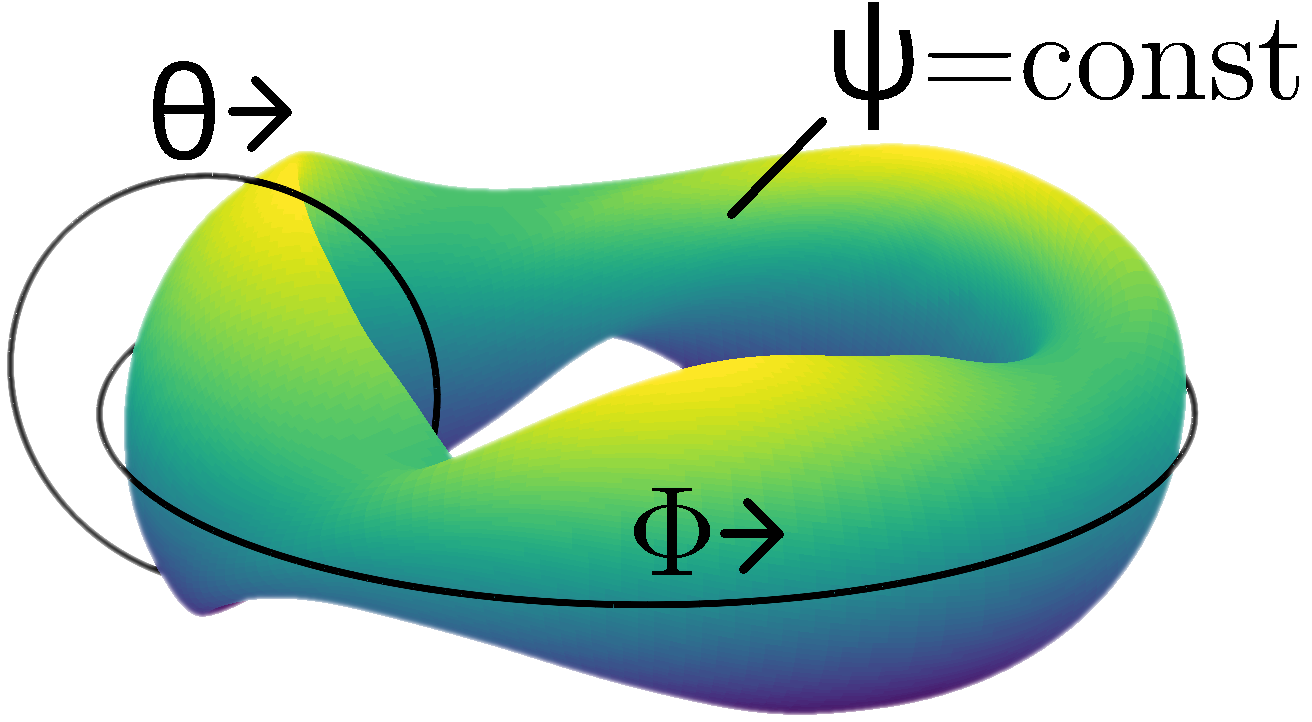}
    \caption{An illustration of a toroidal surface with toroidal angle ($\Phi$), poloidal angle ($\theta$) and flux-surface label $\psi$. Figure adapted from Ref.~\citep{introductionToStellarators2024}.
}
    \label{fig:toroidalGeom}
\end{figure}

We begin by reviewing some of the known basic properties of QS. Several equivalent \citep{Helander2014,rodriguez2022_thesis,freidberg2014idealMHD,burby2020,rodriguez2020a} mathematical descriptions of QS exist. In this Section, we briefly discuss the most relevant equations and coordinates to describe QS and point out their inter-relationships. Here and throughout the paper, we assume the existence of a set of nested toroidal magnetic surfaces, of the kind illustrated in Fig.~\ref{fig:toroidalGeom}. These assumptions are standard within magnetic confinement theory. For details about the coordinates, see for example Ref.~\citep{haeseleer_flux_coordinates}.

It can be shown that QS implies the existence of a divergence-free vector field $\u$ that satisfies the system \citep{rodriguez2020a}
\begin{subequations}
    \begin{align}
    \dl\cdot \u&=0, \label{eq:divu0}\\
     \u\cdot \dl B&=0, \label{eq:udBis0}\\
     \B\times\u &=\dl \psi. \label{eq:Bxu_gradpsi}
\end{align}
\label{eq:u_relabeling_symm}
\end{subequations}
In \eqref{eq:Bxu_gradpsi}, $\psi$  can be any flux function (i.e.~any single-valued function that satisfies $\vec{B} \cdot \nabla \psi = 0$).
It follows directly from \eqref{eq:Bxu_gradpsi} and the divergence-free nature of $\u,\B$ that $\u\cdot\dl$ and $\B\cdot\dl$ commute, i.e.,
\begin{align}
    [\u\cdot\dl,\B\cdot\dl]=0.
    \label{eq:uB_commutator}
\end{align}
Here, we have used the usual definition of the commutator $[A_1, A_2]\equiv A_1 A_2-A_2 A_1$ for two operators $A_1,A_2$.

The mathematical structure of \eqref{eq:u_relabeling_symm} strongly suggests that $\u$ can be associated with a particle-relabelling symmetry group (\cite{padhye1996relabeling,padhye1996fluid,salmon1982hamilton,salmon1988hamiltonian,webb_Zank2006fluid,webb2018magnetohydrodynamics_book} and references therein) with density replaced by the field strength \citep{rodriguez2022_thesis,weitzner2021steadyflow}. Therefore, understanding the hidden symmetry underlying QS might also be beneficial in understanding Casimirs \citep{hameiri2004complete_set_Casimirs,padhye1996relabeling} in ideal MHD and hydrodynamics.

From equation \eqref{eq:Bxu_gradpsi} it follows that
\begin{align}
    \u = (F(\psi)\B-\B \times \dl \psi)/B^2.
    \label{eq:QS_vector_u}
\end{align}
The condition \eqref{eq:udBis0} then implies
\begin{align}
    \B\times \dl \psi \cdot \dl B = F(\psi) \BDB,
    \label{eq:2term_form}
\end{align}
which is sometimes called the two-term form of QS \cite{rodriguez2020a}. 

For a vacuum field $\B=G_0 \dl \Phi$ \cite{boozer1980guidingcenter,Helander2014},
\begin{align}
    F(\psi)=G_0 q_N, \quad q_N\equiv (\iota -N)^{-1},
    \label{eq:F_psi_vac}
\end{align}
where $\iota$ is the rotational transform, $N$ is the helicity of QS, $2\pi G_0/\mu_0$ is the poloidal current outside the surface and $\Phi$ is a toroidal angle. Moreover, the two-term form of QS (equivalently $\u\cdot \dl B=0$) for vacuum fields leads to a traveling wave (TW) solution for $B$ \citep{sengupta_Paul2021vacuum} of the form
\begin{align}
    |\B|=B(\Phi + (F(\psi)/G_0)\alpha,\psi), \quad F(\psi)/G_0=q_N.
    \label{eq:TW_gen_B}
\end{align}
Thus, the rotational transform and, hence, the magnetic shear enters the definition of QS \eqref{eq:2term_form} in a fundamental way. Furthermore, the TW form of $B$ indicates that in the Boozer coordinate system $(\psi,\vartheta_B,\phi_B)$,
\begin{align}
    \phi_B=\Phi, \quad \vartheta_B = \alpha + (\iota-N)\Phi, \quad |\B|=B(\vartheta_B,\psi)
    \label{eq:Boozer_coord_def},
\end{align}
i.e., the field strength $B$ depends only on a single angle $\vartheta_B$. Thus, QS manifests itself in these coordinates. This is true not just for the vacuum field but also in the presence of currents and finite pressure in the plasma.

Another form of QS is the so-called ``triple-product" form \citep{rodriguez2022_thesis,rodriguez2020a}
\begin{align}
    \dl \psi \times \dl B \cdot \dl \lbr \BDB \rbr=0.
    \label{eq:triple_product}
\end{align}
The two-term and the triple-product form of QS are equivalent under the assumption of irrational $\iota$ and continuity on rational surfaces \citep{Helander2014}. Alternatively, the form
\begin{align}
    \u = (\dl \psi\times \dl B)/\BDB,
    \label{eq:u_contra_form}
\end{align}
is consistent with \eqref{eq:QS_vector_u}, satisfies \eqref{eq:udBis0}, \eqref{eq:Bxu_gradpsi}, while \eqref{eq:divu0} implies \eqref{eq:triple_product}. 

It is useful to use the standard Clebsch coordinate system \cite{Helander2014,haeseleer_flux_coordinates} , $(\psi,\alpha,\ell)$, such that $\B=\dl \psi\times \dl \alpha$ with flux label $\psi$, field-line label $\alpha$, and $\ell$, the arclength along the magnetic field. We impose suitable toroidal cuts since $\alpha$ is a multi-valued function. The triple-product form \eqref{eq:triple_product} now reads
\citep{freidberg2014idealMHD}
\begin{align}
  \frac{\del B}{\del \ell} = f(B,\psi).
  \label{eq:basic_BDB_relation}
\end{align}
where $f$ is an arbitrary function of $\psi$ and $B$. Note that the relation is only valid locally. For $B$ with only one magnetic well, it suffices to introduce branchs cut at the extrema of $B(\ell)$, which effectively makes $f$ depend on $\sigma=\text{sign}(\del_\ell B)$ as an additional parameter.

In toroidal geometry, $B$ is subject to the following periodic boundary condition \citep{Helander2014}:
\begin{align}
    B(\psi,\alpha, \ell)= B(\psi, \alpha, \ell +L(\psi)).
    \label{eq:B_periodicity_condition}
\end{align}
The period of $B$, $L(\psi)$, is called the connection length and is a quantity of great physical interest \cite{PhysRevLett.107.115003}. 

As shown in Appendix \ref{app:H}, in $(\psi,\alpha,\ell)$ coordinates, the $\u\cdot\dl B=0$ condition takes the form \cite{freidberg2014idealMHD}
\begin{align}
      \del_\alpha B + H(\psi,\alpha)\del_\ell B=0,
    \label{eq:fund_B_alpha_B_ell_reln}
\end{align}
where $H(\psi,\alpha)$ is independent of $\ell$ and 
depends on the definition of the origin of $\ell$. It follows from \eqref{eq:fund_B_alpha_B_ell_reln} that
\begin{align}
    H(\psi,\alpha)= -\frac{\at{\del_\alpha B}{\ell}}{\at{\del_\ell B}{\alpha}} = \at{\lbr \frac{\del \ell}{\del \alpha}\rbr}{B}
    \label{eq:H_del_ell_del_alpha}
\end{align}
We note that there is a degree of freedom in choosing the $(\ell,\alpha)$ pair. By transforming to the coordinates $(\overline{\ell},\alpha)$, where 
\begin{align}
    \overline{\ell}\equiv \ell -\int H(\psi,\alpha )d\alpha,
    \label{eq:lbar_def}
\end{align}
we find that
\begin{align}
    \u\cdot\dl = -\at{\del_\alpha}{\overline{\ell}}, \;\; \BD=B\del_{\overline{\ell}},\;\;
    \label{eq:uD_BD}
\end{align}
The condition \eqref{eq:fund_B_alpha_B_ell_reln} then reduces to
\begin{align}
    \at{\del_\alpha}{\overline{\ell}}B=0 \quad \text{i.e.}\quad B=B(\overline{\ell},\psi). 
    \label{eq:dalpha_B_TW_B}
\end{align}
Thus, there exists a special TW frame ($\overline{\ell},\alpha$) for QS where the field strength $B$ is independent of $\alpha$ \citep{freidberg2014idealMHD}. We show the equivalence of the two TW forms of $B$ given by \eqref{eq:Boozer_coord_def} and \eqref{eq:dalpha_B_TW_B} in Appendix \ref{sec:equivalence_2term}.

There are, in addition, important integral constraints that follow from the requirements of QS. From the $\del_\ell B$ condition \eqref{eq:basic_BDB_relation}, any integral along the magnetic field line between two points of equal values of $B=B_b$ satisfies
\begin{align}
    \del_\alpha \int_{B\leq B_{b}} d\ell\; F(B;\psi) =0.
    \label{eq:integral_inv_QS_omni}
\end{align}
The field-line independence of the second adiabatic invariant $\Jpl$ 
\begin{align}
\Jpl = \oint \sqrt{\cE-B}\;d\ell,
\label{eq:Jpl}
\end{align}
where $\cE$ is related to the particle energy, is a special case of \eqref{eq:integral_inv_QS_omni}. It follows from \eqref{eq:integral_inv_QS_omni} that quasisymmetric $B$ cannot have local maxima or minima on a flux surface \citep{Helander2014, landremanCatto2012omnigenity}. 

The distinctions between the cases of exact symmetry (axisymmetry or helical symmetry in a straight cylinder), QS and omnigeneity must now be made clear.
For the special cases of exact symmetry, the equations \eqref{eq:u_relabeling_symm} can be solved by letting $u$ be a Killing vector of Euclidean space \citep{burby2020}, which decouples $\u$ from $\B$ such that $\B$ is determined by $\u$. In other words, $u$ is given by the symmetry of the underlying space and does not need to be solved for.
Therefore, the condition \eqref{eq:udBis0}, $\uD B=0$, can be trivially satisfied. Here, we focus on the case of 3D QS where $\u$ cannot be posited \textit{a priori}. Consequently, the condition \eqref{eq:udBis0} imposes a nontrivial constraint. In particular, unlike the case of exact symmetry, the functional form of $f(B,\psi)$ in the $\del_\ell B$ equation must be constrained. We address this question directly in this work. 

QS being a special case of omnigeneity \citep{Helander2014}, the integral property \eqref{eq:integral_inv_QS_omni} also holds for omnigeneous magnetic fields. However, significant differences arise because omnigeneity is not a local property, and a symmetry vector $\u$ does not exist in general for omnigeneous magnetic fields. Thus, the local conditions such as \eqref{eq:2term_form} or \eqref{eq:basic_BDB_relation}, which follow from the properties of $\u$ do not hold for omnigeneity. It follows that omnigenous fields need not be of the TW form \eqref{eq:TW_gen_B}. Furthermore, unlike QS \citep{garrenboozer1991a,landreman2018a}, omnigeneous $B$ need not be uniformly analytic on a flux surface \citep{caryShasharina1997helical,caryShasharina1997omnigenity,plunk2019}.

As a final step, one must impose a force-balance condition. Without imposing force balance, one can satisfy the QS condition exactly order by order in the NAE \citep{rodriguez2021weak}. However, the nonlinear overdetermination problem that results when the force-balance condition is imposed is severely restrictive \citep{garrenboozer1991a,garrenboozer1991b} and analytically prohibitive beyond standard near-axis expansions. This will be addressed in future work.

\section{The connection of QS to integrable PDEs: Painlev\'e property}

In this Section, we take a non-perturbative route to show that the QS-KdV connection can be established using only three fundamental properties of $B$ as discussed in Section \ref{sec:basic_QS}. First, physical quantities such as $B,\psi$ are assumed to be analytic and single-valued functions of their arguments. Second, $B$ must satisfy the periodic boundary condition \eqref{eq:B_periodicity_condition} in $\ell$ on a flux surface with a period $L(\psi)$. Third, $\u$ must be such that a nontrivial special frame exists where $B$ is $\alpha$ independent.

We note that the first assumption is physically reasonable, has been shown to hold near the axis \citep{garrenboozer1991a}, and is also borne out by extensive computations. The second assumption is an essential requirement for QS. (None of the three assumptions hold for omnigeneity.) Regarding the third assumption, an alternative is to demand that $B$ must satisfy the fundamental TW condition \eqref{eq:TW_gen_B} everywhere on the flux surface. The third condition leads to an infinite number of field-line $\alpha$-independent integrals of the form \eqref{eq:integral_inv_QS_omni}.

Before we proceed with our analysis, let us discuss some of the important known exceptional cases of QS. As stated earlier, the case with an exact symmetry is a special case of QS because $\u$ is determined independently of $\B$. Therefore, the third condition $\u\cdot\dl B=0$ is satisfied identically for every choice of $f(B,\psi)$ in the $\del_\ell B$ equation \eqref{eq:basic_BDB_relation}. We find a similar degeneracy in approximate QS near axisymmetry \citep{sengupta_nikulsin_gaur2023QSHBS,plunk2018,plunk2020NASE}, where no constraints exist on $\del_\ell B$. A more general analysis shows that in configurations that are not close to either axisymmetry or helical symmetry, the degeneracy is avoided, and $\del_\ell B$ is once again constrained \citep{nikulsin_sengupta_Jorge_Bhattacharjee_2024asymptotic_GS}. We shall also exclude the exceptional case of zero rotational transform \citep{satoN2022existenceNature} since we focus only on irrational surfaces while assuming continuity on rational surfaces. We shall return to the comparison with the axisymmetric case at the end of this Section.

Our main goal is to study the single-valuedness of $B$ on a constant flux surface $\psi$ that satisfies the $\del_\ell B$ equation \eqref{eq:basic_BDB_relation} and the $\del_\alpha B$ equation \eqref{eq:fund_B_alpha_B_ell_reln}, subject to the periodicity condition \eqref{eq:B_periodicity_condition}.

We first shift to the TW frame where $B$ is $\alpha$ independent, and the $\del_\ell B$ equation can be treated as an ordinary differential equation (ODE) in $\ell$ with period $L(\psi)$. Since $B$ is assumed to be a real analytic function of $\ell$ on the real interval $(0, L)$ with a nonzero radius of convergence, we can extend it to a complex analytic function $\cB(\cZ)$ inside the disk $|\cZ|<L$ with $\cZ$ denoting the complex $\ell$. To be precise, we shall now demand that $\cB$ be a single-valued complex function of $\cZ$, such that it has the same value at every point independent of the path along which it is reached by analytic continuation. 

To proceed further, we need critical ideas from the well-developed theory of complex ODEs \citep{ince1956_ODE,hille1997_Complex_ODE,conte2012painleve_1century}. In contrast to solutions of linear ODEs, whose singularities are determined by those of the coefficients, nonlinear ODEs allow \textit{movable} singularities, which depend on the initial conditions. The standard example is $y'+y^{2}=0$, whose general solution, $y=(x-x_0)^{-1}$, depends on the integration constant $x_0$ determined by the initial condition $y(0)=-x_0^{-1}$. The possible singularities of ODEs are known. They include poles, branch points, essential singular points, and essential singular lines. Single-valuedness of functions breaks down near \textit{critical singularities}, such as branch points of algebraic or logarithmic nature. If the critical singularities are fixed, the functions can be made single-valued by defining suitable branch cuts or fixed Riemann surfaces. However, the presence of singularities, which are both critical and movable, leads to dense multi-valuedness around movable singularities
of solutions and nonintegrability (\cite{conte2012painleve_1century,kruskal2007_Painleve} and references therein). The absence of movable critical singularities in the general solution of an ODE is called the \textit{Painlev\'e property} (PP) of an ODE. 

We now look for $\cB$ with the PP such that there are no critical singularities in the entire complex plane since their very presence can affect $\cB$ on the real line. We now recall some classical results on ODEs with the PP (details given in \citep{ince1956_ODE,hille1997_Complex_ODE,conte2012painleve_1century}). Let us rewrite the $\del_\ell B$ equation in the complex plane as a first-order ODE
\begin{align}
    \cB'(\cZ)=\cF(\cB(\cZ)).
    \label{eq:complex_dBdl}
\end{align}
The classification of the ODEs according to the PP depends on the degree of the ODE, or equivalently the algebraic power of the $\cB'(\cZ)$. The only first-order ODE of the first degree that possesses the PP is the Riccati equation, where $\cF$ is quadratic in $\cB$.

For first-order ODEs of degree $n>1$, the equations of the form 
\begin{align}
    \cB'(\cZ)^n=P_m(\cB(\cZ))
    \label{eq:complex_Poly_dBdl}
\end{align}
with $P_m$ a polynomial of degree $m$, possess the PP provided $n\leq 2m$. Furthermore, only four possibilities exhaust all ODEs of the form \eqref{eq:complex_Poly_dBdl}. With suitable variable transformations, all of them can be reduced to the following ODE with a cubic nonlinearity
\begin{align}
    \cB'(\cZ)^2=P_3(\cB(\cZ)),
    \label{eq:complex_cubic_dBdl}
\end{align}
which is exactly solvable in terms of elliptic functions and their degenerate cases. As is well-known, elliptic functions are doubly-periodic, meromorphic functions. The dependence on the arbitrary initial condition or the TW frame thus leads to only movable poles and no movable critical singularities. Hence, the solution possesses the PP. These classical results follow from the standard Painlev\'e test and extensions of Malmquist's theorem \citep{hille1997_Complex_ODE,eremenko1982meromorphic}. They show how stringent and rigid the constraint of single-valuedness can be.

Returning to the QS problem, we first observe that the Riccati equation for $B$ is inconsistent with the NAE solution. The Riccati equation for $B$ has constant real coefficients on a flux surface, which does not admit nontrivial periodic solutions on the real line \cite{coddington_levinson1956_ODEtheory}. 
For degrees $n>1$, the periodicity condition can be easily satisfied with the elliptic functions for the cubic and quartic cases. Now, the cubic and quartic polynomials are the twice-integrated forms of the KdV and Gardner's equations. The PP of the KdV and Gardner's equation in the TW frame is well-known and is conjectured to be essential for integrable PDEs with soliton solutions \citep{ARS_I, ARS_II}. 

We now address the question: is the PP necessary and sufficient for a quasisymmetric $B$? It is certainly sufficient for QS since both the triple product form \eqref{eq:triple_product} and the periodicity condition can be satisfied. However, the necessity is not apparent, particularly in light of axisymmetry, where no such stringent restriction on $\del_\ell B$ appears. In axisymmetry, weak singularities such as $R^m\log{R}^n$, with integer $m,n$, often arise in the description of $\psi$ \citep{cerfon2010one_size} and in $B$ in cylindrical coordinates. However, these singularities are fixed and, therefore, not of the critical movable type. To further understand the difference between a perfectly axisymmetric system and a system with excellent but not exact QS, we need to look deeper into the $\u\cdot\dl B=0$ condition. As discussed in Appendix \ref{app:H}, if QS were exact, we would have the same characteristics as axisymmetry. However, unlike the axisymmetric problem, the QS problem is intrinsically overdetermined and three-dimensional. With this in mind, we now motivate the PP as a criterion that allows one to look for robust quasisymmetric systems. 

In axisymmetry, all field lines are equivalent, and the TW frame, where $H=0$, is the same for all $\alpha$. Regarding the effect on the singularities, this implies that the dependence of $B$ on the field line label is precisely the same for each singularity, which amounts to a trivial rigid shift of the frame. However, in a generic approximate 3D QS, the geometry and $B$ are strongly coupled. Due to the overdetermined nature, the $\u\cdot\dl B=0$ condition can only be approximately satisfied. As a consequence, the TW frame $H=0$ is, in general, different for different field lines. Thus, the $\alpha$ dependence of the singularities of $\del_\ell B$ varies from field line to field line. Since these movable singularities can be critical, single-valuedness of $B$ is harder to achieve. However, when we demand the PP, the moving singularities are not critical, which respects the single-valuedness of $B$. As shown in Sections \ref{sec:numerical_verification} and \ref{sec:Data_Driven}, every numerically optimized quasisymmetric stellarator has the PP. Furthermore, as the optimization progresses, the PP emerges quickly and is maintained. Therefore, we argue that the PP is necessary and sufficient for obtaining robust QS in a 3D stellarator using numerical optimization methods, where single-valuedness and periodicity of $B$ are enforced directly using a Fourier representation.

Finally, we note that none of the above arguments required knowledge of the exact force balance condition. The roots of the polynomial in \eqref{eq:complex_cubic_dBdl} that determine the zeros of $(\del_\ell B)$ can vary with the flux surface label, and force balance determines their profiles.
We explore the behavior of the roots of $(\del_\ell B)^2$ in the next section, in the context of numerically optimized quasisymmetric stellarator equilibria.

\section{Reduced dimensionality of quasisymmetric $B$}
\label{sec:reduced_dim}

\begin{figure}
    \centering
    \includegraphics[width=.45\textwidth]{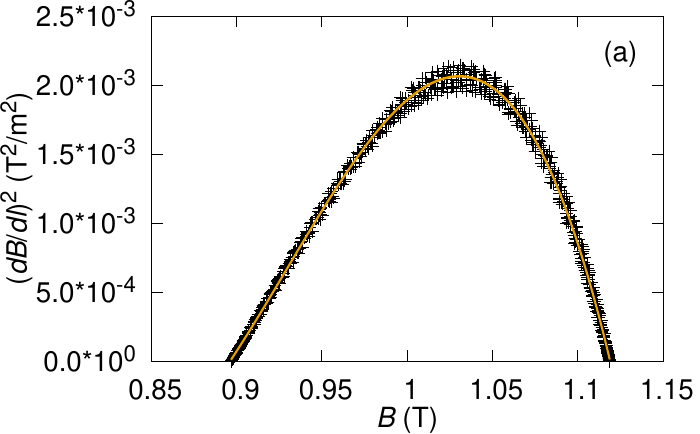 }
    \includegraphics[width=.45\textwidth]{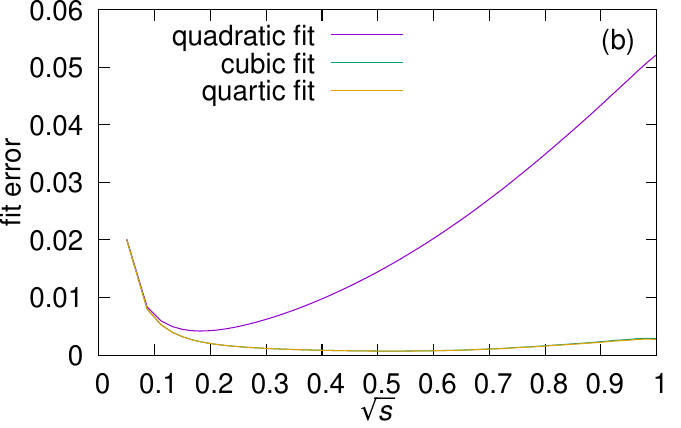 }
    \caption{The value of $(dB/d\ell)^2$ on the outermost flux surface in the precise QA equilibrium \cite{landreman2021} is plotted as a function of $B$, together with a cubic polynomial fit in (a). In (b), the errors of quadratic, cubic, and quartic fits are compared on each flux surface, demonstrating that a cubic ansatz (as predicted by KdV) is sufficient. Hence, precise QA has the Painlev\'e property for $B$.}
   \label{fig:ansatz_validation}
\end{figure}

\begin{figure}
    \centering
    \includegraphics[width=.41\textwidth]{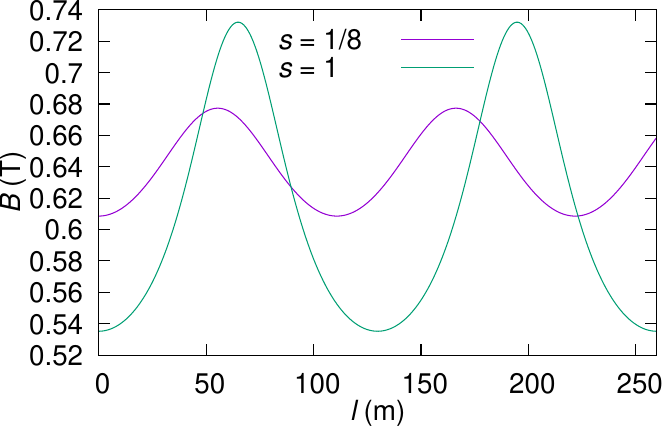 }
    \caption{The value of $B$ in a quasiaxisymmetric device in the core ($s=1/8$) and at the edge ($s=1$). The separatrix is outside the device but close to its edge. 
    Note the increasing period as the separatrix is approached.}
     \label{fig:soliton}
\end{figure}
 
\begin{figure}
    \centering
    \includegraphics[width=.4\textwidth]{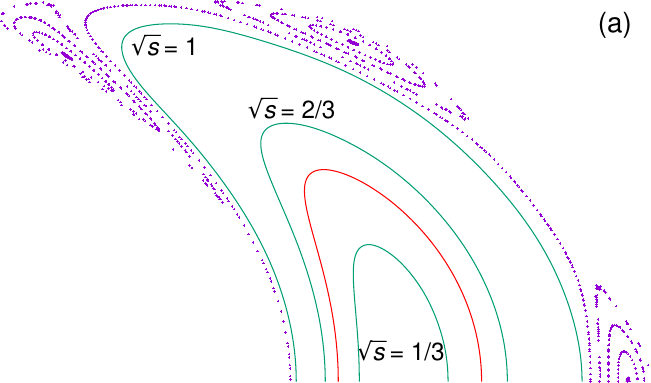 }    \includegraphics[width=.4\textwidth]{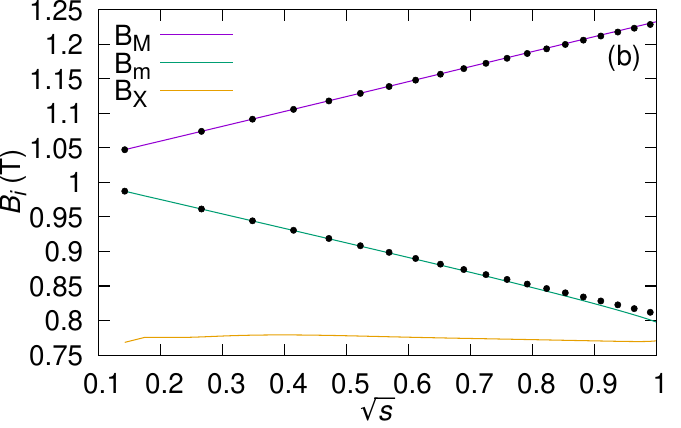 }
    \caption{Prediction of the maximum toroidal volume that can be made quasisymmetric for precise QA. The boundary of the extended precise QA and two other flux surfaces are shown inside a Poincar\'e plot of the islands in (a), and the roots are plotted in (b). The boundary of the original precise QA is shown in red. Note that the islands occur where the second and third roots approximately touch.}
    \label{fig:XpQA_roots}
\end{figure}

In the previous Section, we have shown that $(\del_\ell B)^2$ is a cubic or quartic polynomial in $B$, whose roots can only be functions of $\psi$. Therefore, the PP significantly reduces the dimensionality of quasisymmetric $B$, which we shall now fully explore. We restrict ourselves to the cubic case,
\begin{align}
    \lbr \frac{\del B}{\del \ell}\rbr^2 = \cD(\psi) (B_M-B)(B- B_m)(B-B_X),
    \label{eq:cubic_ansatz}
\end{align}
where the roots of the cubic polynomial, $B_i=\{B_M,B_m,B_X\}$, are functions of $\psi$, and $\cD(\psi)$ is a proportionality factor. Since $\BDB$ vanishes at the roots, the functions $B_M$ and $B_m$ denote the maximum and minimum values of $B$ on a flux surface, while $B_X$ is a third root of the polynomial, possibly a saddle. The exact solution of \eqref{eq:cubic_ansatz}, subject to the periodicity constraint \eqref{eq:B_periodicity_condition}, can be expressed in terms of elliptic functions \citep{kamchatnov2000nonlinear,novikov1984theory_solitons} as
\begin{align}
  B(\ell) &= B_X + (B_M - B_X)\,
            \mathrm{dn}^2\!\!\left(
              \sqrt{\frac{\cD}{2}\frac{(B_M-B_X)}{2}}\,\ell,\,m
            \right),\nonumber\\
            m &= (B_M - B_m)/(B_M - B_X), \;\; 0\leq m \leq 1,
  \label{eq:dn_solution}
\end{align}
where $\mathrm{dn}(w,m)$ is the Jacobi delta-amplitude elliptic
function. Note that, due to translational symmetry in $\alpha$, we can replace $\ell$ by $\overline{\ell}$ given by \eqref{eq:lbar_def}. 

Equation~\eqref{eq:cubic_ansatz} and its solution contain four functions of $\psi$: the three
roots $\{B_M, B_m, B_X\}$ where $\BDB$ vanish, and the prefactor $\mathcal{D}(\psi)$. We now show that, if flux surfaces are nested and Boozer coordinates can be defined, $\mathcal{D}$ is not an independent free function but is fully determined by the roots and the rotational
transform $\iota(\psi)$, reducing the description of $B$ on a flux
surface to exactly three spectral parameters. We describe our reasoning below.

Since $B$ is periodic with period $L(\psi)$, integrating the
field-line equation in Boozer coordinates over one full period gives a
net change of $2\pi$ in the Boozer angle $\vartheta$
\begin{equation}
  \int_0^{L} \frac{\partial\vartheta}{\partial\ell}\,d\ell = 2\pi.
  \label{eq:period_boozer}
\end{equation}
Using $\BD\vartheta=B\del_\ell \vartheta$ and the Boozer Jacobian relation $\BD\vartheta=\iota_N B^2/(G+\iota I)$ \citep{Helander2014}, we obtain
\begin{equation}
  \oint B\,d\ell = \frac{2\pi(G + \iota I)}{\iota_N},
  \label{eq:Bavg_constraint}
\end{equation}
where $\iota_N\equiv \iota -N$, and $G(\psi)$ and $I(\psi)$ are the poloidal and toroidal current
flux functions. For purely vacuum fields $I=0$, but we keep the general expressions here. 

Integrating over one period using the identity
$\int_0^{K(m)}\mathrm{dn}^2(w,m)\,dw = E(m)$ yields
\begin{equation}
  \oint B\,d\ell
  = L \lbr B_X + (B_M - B_X)\frac{E(m)}{K(m)}\rbr,
  \label{eq:Bavg_elliptic}
\end{equation}
where $K(m)$ and $E(m)$ are the complete elliptic integrals of the
first and second kind. The period $L$ obtained from \eqref{eq:Bavg_constraint} and \eqref{eq:Bavg_elliptic} reads
\begin{align}
    L(\psi)=\frac{2\pi(G+\iota I)}{\iota_N}\lbr B_X + (B_M - B_X)\frac{E(m)}{K(m)}\rbr^{-1}.\label{eq:BigL}
\end{align}

The quantities $L$ and $\mathcal{D}$ are related through the periodicity of the
$\mathrm{dn}$ function by
\begin{equation}
  \sqrt{\mathcal{D}}\,L(\psi)
  = \frac{4K(m)}{\sqrt{B_M - B_X}}.
  \label{eq:DL_relation}
\end{equation}
Substituting Eqs.~\eqref{eq:Bavg_elliptic} and~\eqref{eq:DL_relation}
into the geometric constraint~\eqref{eq:Bavg_constraint} and solving
for $\sqrt{\mathcal{D}}$, we obtain
\begin{equation}
  \sqrt{\mathcal{D}}
  = \frac{\iota_N}{G + \iota I}
    \lbr \frac{K}{\pi/2}\rbr
    \frac{B_X + (B_M-B_X)\,E/K}{\sqrt{B_M - B_X}}.
\label{eq:D_formula}
\end{equation}
Every factor on the right-hand side of Eq.~\eqref{eq:D_formula} is
either a flux function ($G$, $I$, $\iota$) or depends on
$\psi$ only through the three roots, since $m$, $K(m)$, and $E(m)$
are all determined by $\{B_M, B_m, B_X\}$ alone. Therefore
\begin{equation}
  \mathcal{D} = \mathcal{D}\!\bigl(B_M(\psi),\,B_m(\psi),\,
                B_X(\psi),\,\iota(\psi)\bigr),
  \label{eq:D_dependence}
\end{equation}
and the description of quasisymmetric $B$ on a flux surface requires
exactly three spectral functions of $\psi$ (the three roots of the cubic)
plus the rotational transform profile, which is constrained
independently by force balance. This is the precise statement of the
hidden lower dimensionality of $B$.


We now show that the cubic formula for $(\del_\ell B)^2$ given in \eqref{eq:cubic_ansatz} is connected to the KdV equation,
\begin{align}
\del_t B+6 B \del_x B+ \del_x^3 B=0.
\label{eq:modB_KdV}
\end{align}
We differentiate \eqref{eq:cubic_ansatz} twice with respect to $\ell$ and use the $\del_\alpha B$ equation \eqref{eq:fund_B_alpha_B_ell_reln} to obtain the KdV equation for $B$. We find that the coordinates $(x,t)$ are related to $(\alpha,\ell)$ through
\begin{align}
    \frac{x}{\ell}=\sqrt{\frac{\cD}{2}},\;\; \frac{t}{\int d\alpha H} =  \frac{\sqrt{\cD/2}}{2(B_M+B_m+B_X)}.
    \label{eq:x_t_def}
\end{align}

The $\del_\alpha B$ relation \eqref{eq:fund_B_alpha_B_ell_reln}, now takes the following form of a 1D TW equation with speed  $c(\psi)$,
\begin{align}
    \frac{\del B}{\del t} + c(\psi)\frac{\del B}{\del x} =0, \quad c(\psi)= 2(B_M+B_m+B_X).
    \label{eq:fund_TW_modB}
\end{align}
Since the speed $c(\psi)$ is constant on a flux surface, we reiterate that quasisymmetric $B$ satisfying \eqref{eq:modB_KdV} must be periodic TW solutions of the KdV equation. We show this explicitly by rewriting the exact solution for $B$ \eqref{eq:dn_solution} in terms of the cnoidal solution \citep{kamchatnov2000nonlinear} of KdV,
\begin{align}
    B &=B_m +(B_M-B_m)\text{cn}^2\lbr \sqrt{\frac{B_M-B_X}{2}}(x-c t),m\rbr.
    \label{eq:global_B_exp}
\end{align}

Alternatively, we can start with the KdV equation, move to the TW frame, and integrate the KdV equation twice to recover \eqref{eq:cubic_ansatz}. The adiabatic invariant and the integrals \eqref{eq:integral_inv_QS_omni} obtained from KdV potentials are manifestly time ($\alpha$) independent \citep{gjaja_Bhattacharjee_1992asymptotics_reflectionless,berry_Howles_1990fake_Airy_reflectionless,brezhnev2008_integrabiliy_FG} because KdV shares the Painlev\'e property.

The Jacobi cosine elliptic function $\text{cn}(u,m)$ with modulus $m$  has maximum amplitude of unity. Thus, \eqref{eq:global_B_exp} confirms that $B$ varies between the minimum and the maximum values $B_m,B_M$ of $B$. Two limiting cases, $m \to 0$ and $m \to 1$, are of particular interest. In the $m\to 0$ limit, $\mathrm{cn}(u,m)\to \cos(u)$ whereas in the $m\to 1$ limit $\mathrm{cn}(u,m)\to \mathrm{sech}(u)$. Therefore, $B$ is sinusoidal in the first limit and the KdV reflectionless potential \citep{gjaja_Bhattacharjee_1992asymptotics_reflectionless} in the second limit. 

In the $m\to 0$ limit, both $B_m$ and $B_M$ approach a common value (say $B_0$) from \eqref{eq:dn_solution} and $B$ is sinusoidal. This is the NAE limit, where, the on the axis $B$ is a constant, $B_0$, and varies as $B\approx B_0(1 + \etab \sqrt{2\psi} \cos\vartheta_B) + 2\psi(B_{20} + B_{22}\cos 2\vartheta_B)$. The constant $\etab$ dominates the maximum variation of $B$ on a flux surface, with the $O(2\psi)$ term being only a small second-order correction. Thus, the well-understood behavior of $B$ in Boozer coordinates near the magnetic axis \citep{garrenboozer1991a,garrenboozer1991b,landreman2018a} is consistent with \eqref{eq:cubic_ansatz}. As $m\to 0$, $E(m),K(m)$ both approach $\pi/2$, which implies that $L\to 2\pi (G+\iota I)/(B_0 \iota_N)$ and $\sqrt{D}\to \iota_N B_0/( (G+\iota I)\sqrt{B_0-B_X})$ are both finite provided $B_0-B_X>0$ and $\iota_N \neq 0$. Further, the NAE predictions are
\begin{align}
    \frac{B_M}{B_0},\frac{B_m}{B_0}=1\pm \etab \sqrt{2\psi},\quad \frac{B_X}{B_0} = \frac{4 B_{22}/B_0 + \etab^2}{4B_{22}/B_0 + 2\etab^2}.
    \label{eq:NAE_predictions}
\end{align}
Since $B_{22},B_0,\etab$ are all constants related to the $B$ spectrum in Boozer coordinates, within the NAE framework, $B_X$ and $\cD$ are constant throughout the volume. 

We now proceed to the $m\to 1$ limit. We caution that this limit must be interpreted with care.
In this limit, $E\to 1$, but $K(m)\to \ln{4/\sqrt{1-m}}$ diverges logarithmically, and $B$ approaches a soliton profile. While the cnoidal solution \eqref{eq:dn_solution} describes periodic functions with finite connection length, the soliton profile is non-periodic and corresponds to an infinite domain. Thus, this limit represents a breakdown of the periodicity assumption underlying QS in toroidal geometry. This has important consequences. Expressions for $L$ and $\sqrt{D}$ in \eqref{eq:BigL} and \eqref{eq:D_formula} were obtained under the assumption that $B$ remains periodic with finite $L$, and therefore cannot be naively extrapolated to the $m\to 1$ limit, where $\sqrt{D}L$ diverges. The apparent finiteness of $L$ obtained by direct substitution reflects the use of formulae outside their domain of validity. Physically, the degeneration to a solitary-wave profile signals the loss of global periodicity and the breakdown of Boozer straight-field-line coordinates. In this regime, the connection length should be regarded as diverging or ceasing to be well-defined in the periodic sense. Since the non-periodic $m\to 1$ limit may not be correctly captured in our calculations, it is instructive to think about physical situations where periodicity is broken in stellarators.

Firstly, the limit in which $\iota_N \to 0$ corresponds to standard X-points in tokamaks, which are well understood. In this case, $L_c \sim 1/\iota$, and periodicity along a field line is lost because the field line cannot cross the X-points. 
This scaling is in fact correctly captured by \eqref{eq:BigL}, where, if $\iota_N$ or $B_X$ go to zero, $L$ will diverge while $\sqrt{D}\to 0$, since the divergence in $\sqrt{D}$ only logarithmic. If QS is exact, X-points can form only when $\iota$ resonates with the helicity of QS \citep{rodriguez2021islands}. However, even for minor QS errors, small resonant magnetic fields can lead to the formation of X-points and islands.

Secondly, for irrational $\iota$, sharp structures on the flux surfaces ("ridges") can also arise \citep{bader2019,punjabi2020}. These sharp structures can also hinder magnetic field lines from crossing them \citep{boozer2015}.
In both these cases, if the field lines can not cross the X-points or ridges then they can not physically connect two consecutive maxima or minima of $B$, leading to a divergence of the connection length.

The above physically motivated divergence of the connection length shows how the assumptions used to derive the expressions for $L,\sqrt{D}$ are no longer valid in the $m\to1$ limit. Indeed, equation \eqref{eq:Bavg_constraint} presupposes that $L(\psi)$ is finite, since $B$ has a positive minimum, and the integral of such a function over an infinite domain cannot be finite. Therefore, the loss of periodicity in $B$ corresponds to a breakdown of global Boozer straight-field-line coordinates.

Thus, while not mathematically proven within the context of our calculations, we speculate that the $m\to 1$ limit corresponds to the formation of an X-point or a sharp ridge on the flux surface. In Section \ref{sec:numerical_verification}, we indeed observe the formation of an X-point as we approach the limiting $m\to 1$ flux surface. In practical terms, this means that QS stellarators may naturally form sharp edges as the volume of quasisymmetry is pushed to its limit, e.g., for compact devices. Sharp edges are essential for non-resonant divertors \citep{boozer2015, punjabi2020}. This could potentially have significant beneficial consequences for stellarator optimization, since these divertors would be obtained naturally by optimizing for quasisymmetry in compact devices.

Interestingly, a similar singular limit also appears in the isodynamic limit, where the magnetic field curvature has cnoidal solutions \citep{schief2003nested}. The $m=1$ limit is never achieved because of a geometrical obstruction when the central toroidal hole pinches. A similar pinching effect, leading to formation of cusps and break down of nestedness in quasiaxisymmetric devices, has been discussed in \citep{plunk2020NASE,brownSenguptaNikulsin2025Palumbo}. Other topological obstructions such as the ``Reeb components" \citep{duignan2024global} may arise before the $m\to 1$ limit is obtained. The Reeb case requires the toroidal magnetic field to vanish at some point if MHS force balance is imposed \citep{duignan2024global}. 

\section{Numerical verification}
\label{sec:numerical_verification}

\begin{figure}
    \centering
    \includegraphics[width=0.4\textwidth]{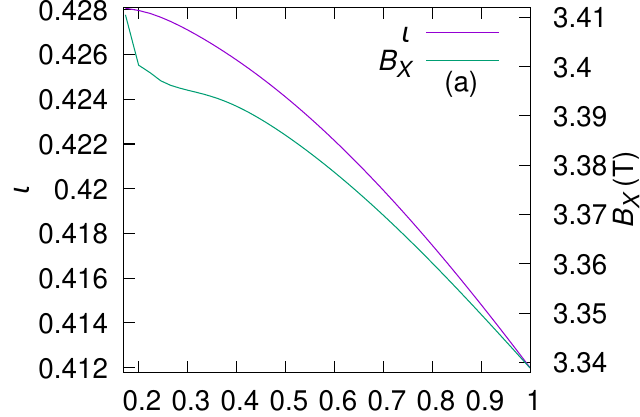 }
    \includegraphics[width=0.4\textwidth]{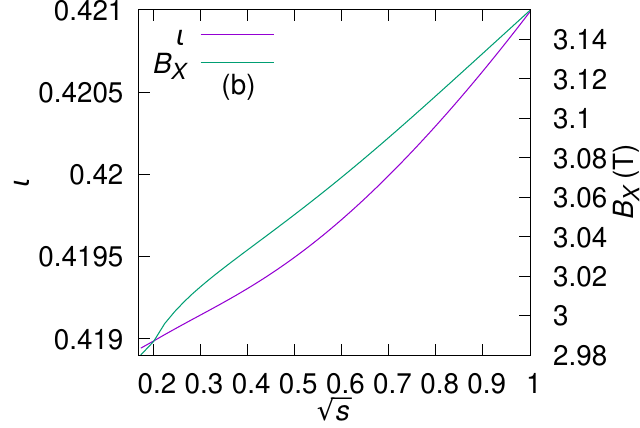 }
    \caption{The $B_X$ and $\iota$ profiles plotted for precise QA versions with negative shear (a) and positive shear (b). Symmetry-breaking modes have been filtered out.
}
    \label{fig:bx_iota}
\end{figure}

\begin{figure*}
    \centering
    \includegraphics[width=\textwidth]{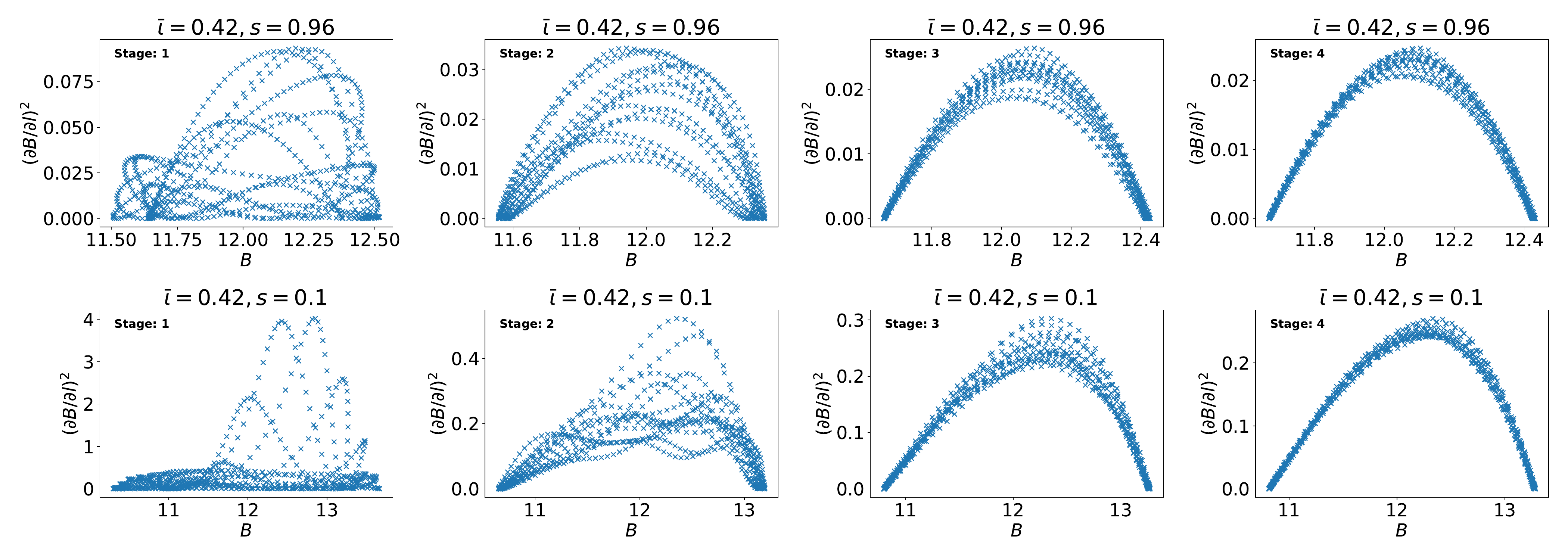}
    \caption{$(\partial_l B)^2$ as a function of $B$, on two surfaces, across the four stages of optimization for the Landreman-Paul precise QA configuration. Top row: surface near edge at $s=0.96$. Bottom row: surface near core at $s=0.1$. Left to right: increasing stages in the Fourier continuation based optimization \cite{landreman2021}.}
    \label{fig:ansatz_validation2}
\end{figure*}

\begin{figure}
    \centering
    \includegraphics[width=0.5\textwidth]{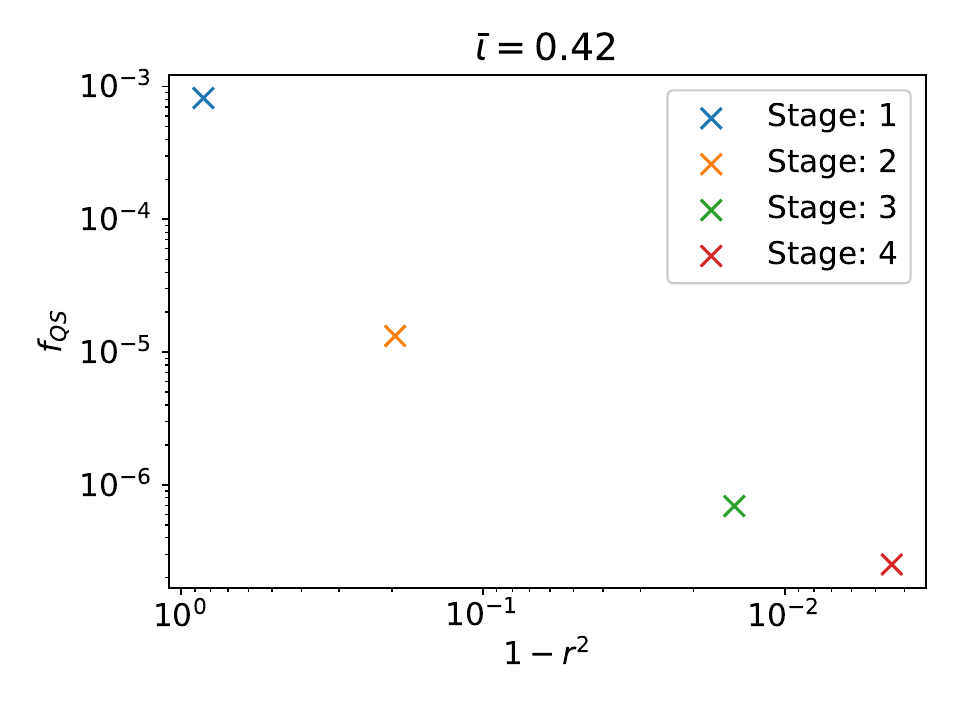}
    \caption{The two-term QS error, summed over the surfaces, is plotted on the ordinate against $1-r^2$ (averaged across all surfaces) on the abscissa, where $r^2$ is the coefficient of determination for a cubic best fit.}
    \label{fig:staging,r2vsQSerror}
\end{figure}

\begin{figure}
    \centering
    \includegraphics[width=0.5\textwidth]{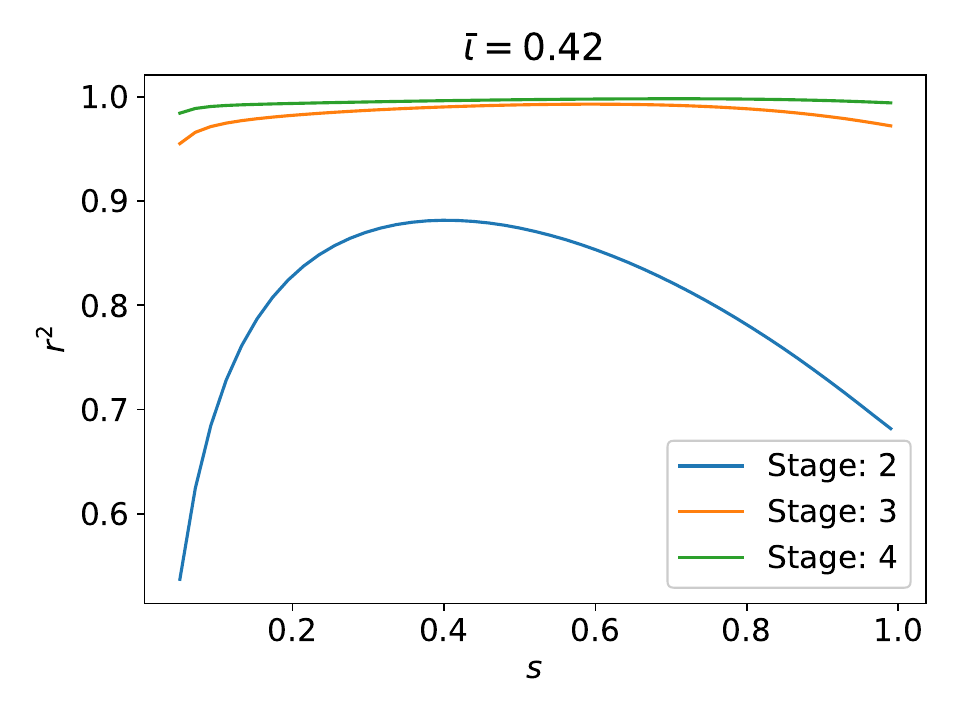}
    \caption{For a cubic best fit, we plot the coefficient of determination against the surfaces of the Landreman-Paul configuration across the latter three stages of the optimization.}
    \label{fig:staging,r2surfs}
\end{figure}

\begin{figure*}
    \centering
    \includegraphics[width=\textwidth]{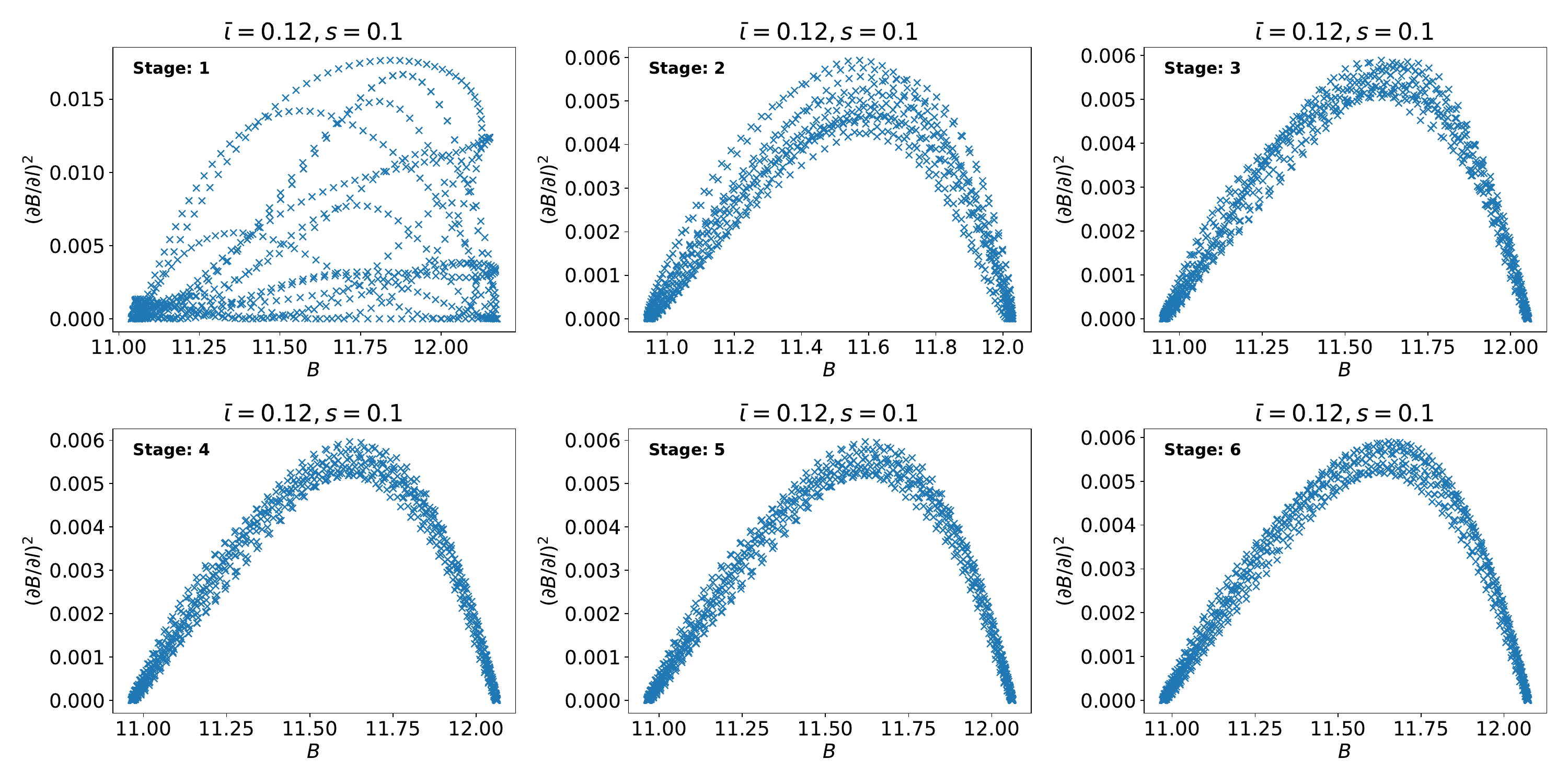}
    \caption{$(\partial_l B)^2$ as a function of $B$, for $s=0.1$, across the six stages of optimization for the $\bar{\iota}=0.12$ Buller configuration.}
    \label{fig:staging,quartic,dBdl^2vsB,s=0.1}
\end{figure*}

\begin{figure*}
    \centering
    \includegraphics[width=\textwidth]{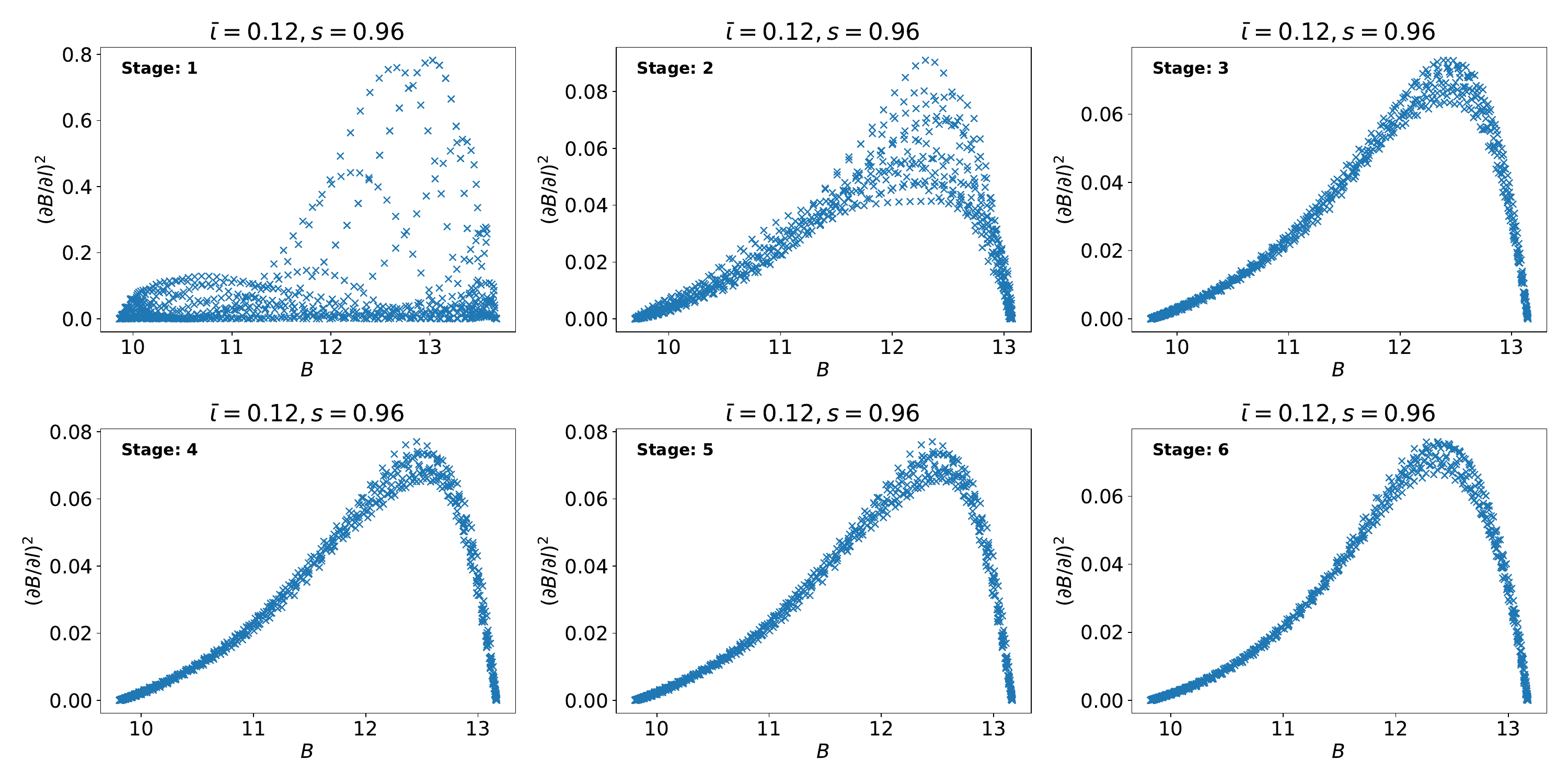}
    \caption{$(\partial_l B)^2$ as a function of $B$, for $s=0.96$, across the six stages of optimization for the $\bar{\iota}=0.12$ Buller configuration.}
    \label{fig:staging,quartic,dBdl^2vsB,s=0.96}
\end{figure*}

\begin{figure}
    \centering
    \includegraphics[width=0.4\textwidth]{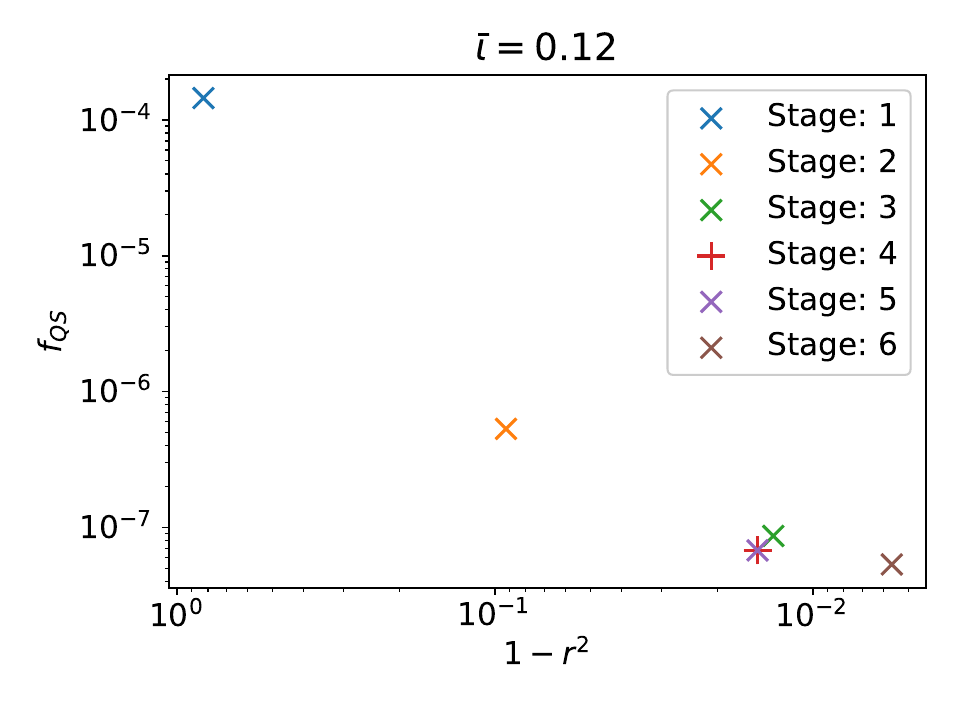}
    \caption{The two-term QS error, summed over the surfaces, is plotted against $1-r^2$ (averaged across all surfaces), where $r^2$ is the coefficient of determination for a quartic best fit.}
    \label{fig:staging,r2vsQSerror,quartic}
\end{figure}

\begin{figure}
    \centering
    \includegraphics[width=0.4\textwidth]{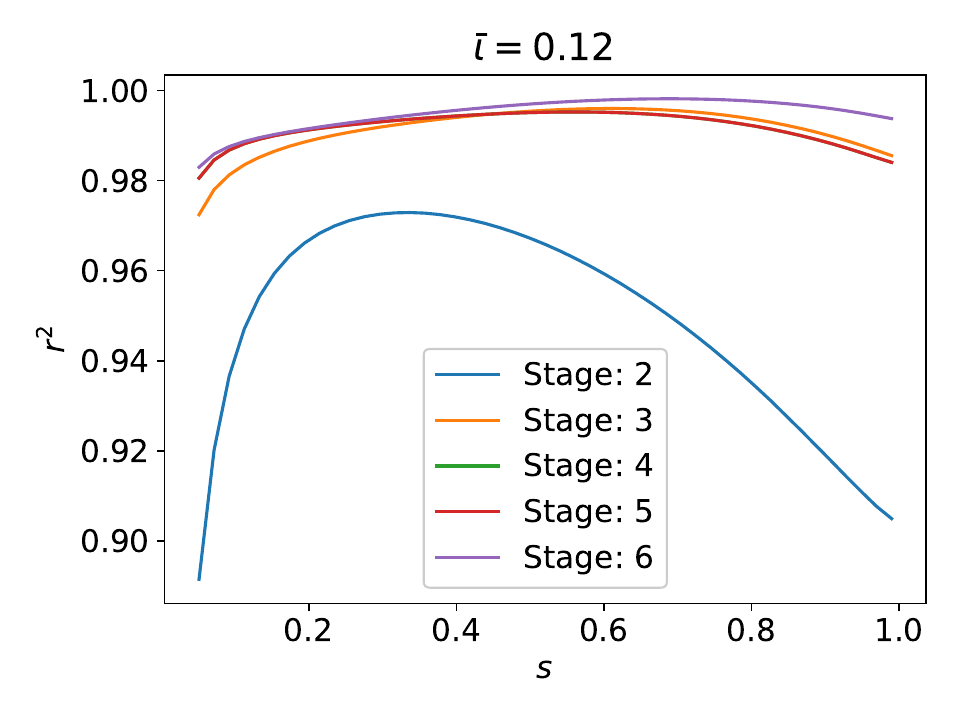}
    \caption{For a quartic best fit, we plot the coefficient of determination against the surfaces of the Landreman-Paul configuration across the latter five stages of the optimization.}
    \label{fig:staging,r2surfs,quartic}
\end{figure}

We now demonstrate numerically that a large class of quasisymmetric $B$ satisfies the integrated form of KdV as given in \eqref{eq:cubic_ansatz}. 
As demonstrated in Fig. \ref{fig:ansatz_validation}(a), $(dB/dl)^2$, when plotted as a function of $B$, has the form of a cubic polynomial for the precise QA equilibrium \cite{landreman2021}; the scatter is due to imperfect QS. On the other hand, Fig. \ref {fig:ansatz_validation}(b) shows that increasing the polynomial order beyond three does not lead to an observable decrease in error (defined as $1-r^2$, where $r^2$ is the coefficient of determination), as the fit error is dominated by QS error. Thus, Fig. \ref{fig:ansatz_validation} supports equation \eqref{eq:cubic_ansatz} as a model for a device with perfect QS. A similar analysis for other configurations, which tend to have greater QS errors, is shown in Appendix \ref{app:additional_fig}. (Note that $s$ is the normalized toroidal flux in all figures.)

With the global description of the exact quasisymmetric $B$, \eqref{eq:global_B_exp}, at our disposal, we now investigate a key element of the theory: the physical meaning of the roots $B_i$. In Fig. \ref{fig:XpQA_roots}(b), we show the radial dependence of the roots. Symmetry-breaking modes have been filtered out, and the roots are obtained by a cubic fit of the numerical data. Points represent the actual minimum and maximum values of $B$ on each flux surface. Surprisingly, the NAE description of $B_M$ and $B_m$ continues to hold even far from the axis.
As shown in Fig. \ref{fig:XpQA_roots}(a), the location where $B_m$ and $B_X$ meet coincides with the location of an X-point that leads to a chain of islands. This is consistent with periodicity breaking down for $m\to 1$.

The third root $B_X$ depends quite sensitively on the $\iota$ profiles (Fig. \ref{fig:bx_iota}).
We show in Fig. \ref{fig:bx_iota} that, for small but finite shear, $B_X$ tracks the $\iota$ profile quite closely but varies more for positive shear.

Lastly, we show how the integrated form of KdV is approached through the stages of optimization, where the parameter space is progressively expanded. In Fig. \ref{fig:ansatz_validation2}, for the Landreman-Paul precise QA configuration, not only does the `noise' decrease (as expected from improving QS), but the shape becomes increasingly cubic. We show this both for a surface close to the axis and a surface close to the edge. Figures \ref{fig:staging,r2vsQSerror} and \ref{fig:staging,r2surfs} further show how the integrated KdV cubic increases in accuracy (as determined by the coefficient of determination) further into the optimization process. In Figures \ref{fig:staging,quartic,dBdl^2vsB,s=0.1}, \ref{fig:staging,quartic,dBdl^2vsB,s=0.96}, \ref{fig:staging,r2vsQSerror,quartic}, and \ref{fig:staging,r2surfs,quartic}, we show similar plots for the low mean rotational transform configurations introduced in \cite{buller2024family}. Note that the trend is towards quartic instead of cubic. As evident from these plots, the PP is established in the entire volume within the very first few stages of optimization and is maintained thereafter. 

\section{Data driven verification}
\label{sec:Data_Driven}

\begin{figure*}
    \centering
    \includegraphics[width=\textwidth]{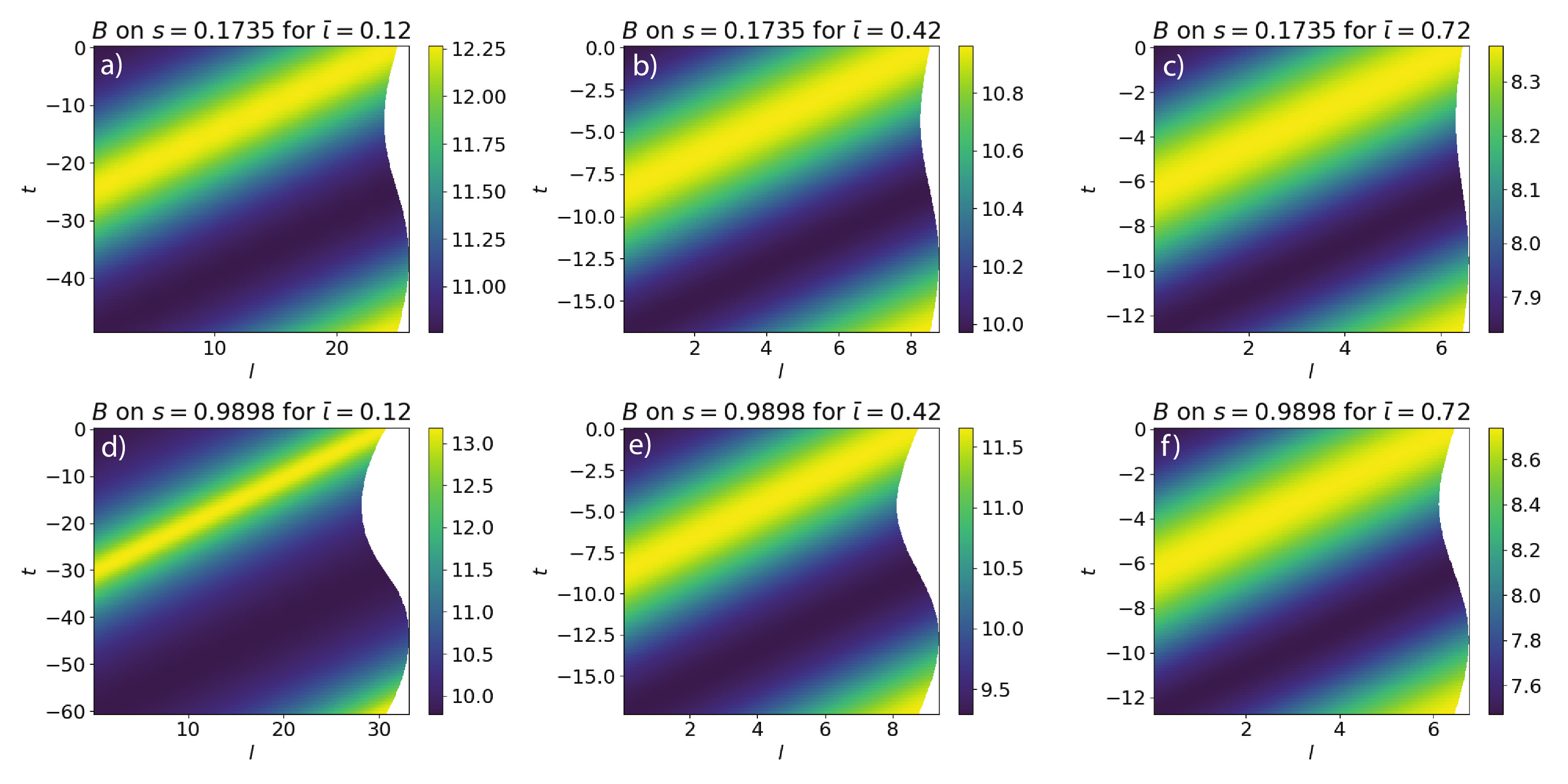}
    \caption{$B$ on a magnetic surface parameterised by $l$ and $t$, for precise-QA-like configurations with varying rotational transform\cite{buller2024family}. Top row: flux surface close to the core ($s=0.1735$. Bottom row: flux surface close to the edge $s=0.9898$.}
    \label{fig:traveling wave}
\end{figure*}

\begin{figure*}
   \centering
    \includegraphics[width=1\textwidth]{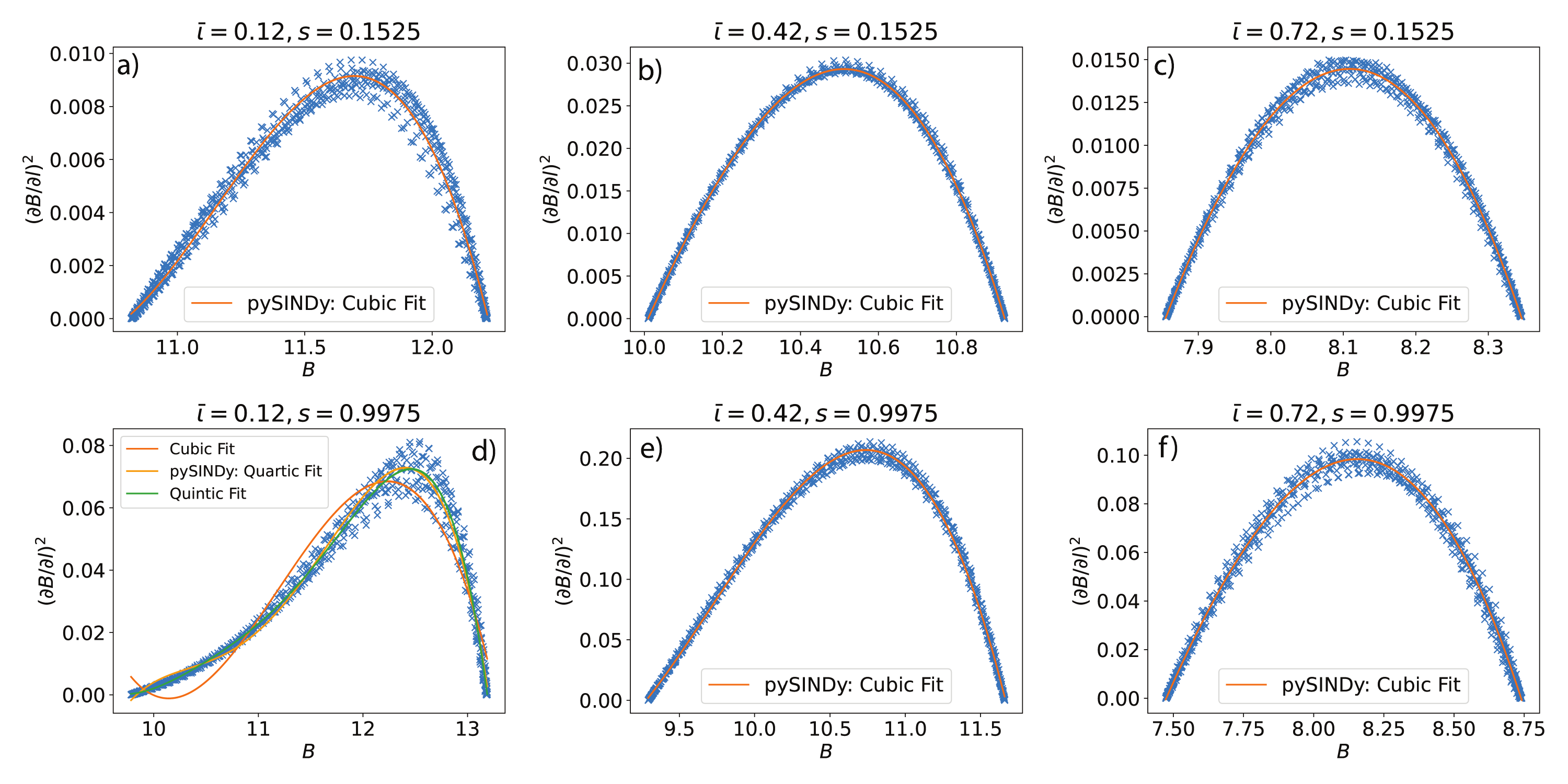}
    \caption{$(\partial_l B)^2$ as a function of $B$ on two surfaces, for precise-QA-like configurations with varying rotational transform\cite{buller2024family}. Only at low rotational transform and far away from the axis, as in d), is a cubic a poor fit.}
    \label{fig:dBdl^2vsB}
\end{figure*}

\begin{figure*}
    \centering
    \includegraphics[width=\textwidth]{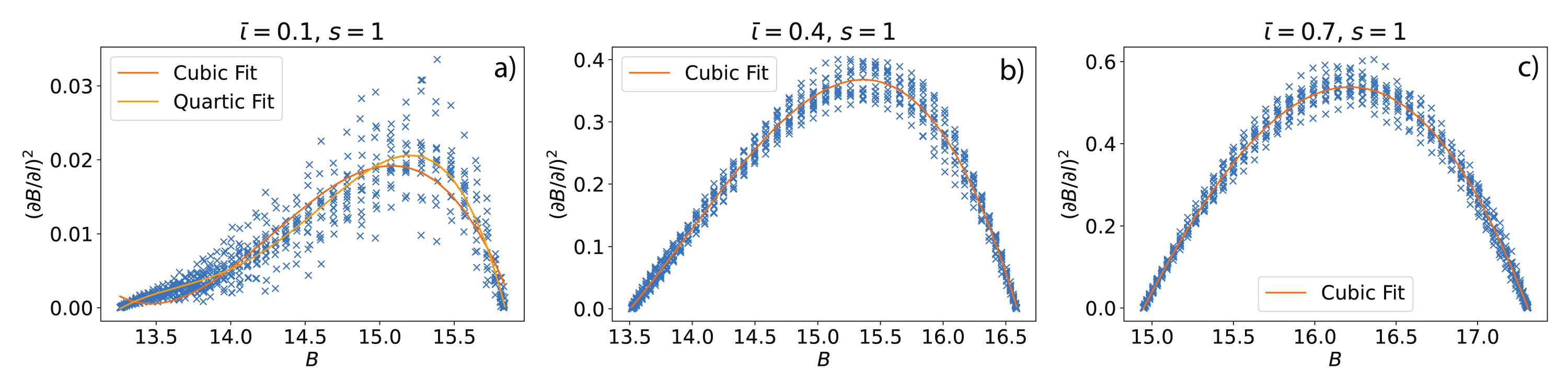}
    \caption{$(\partial_l B)^2$ as a function of $B$ on the outermost surface, on configurations at three different mean rotational transforms selected from the QUASR database. Only at low rotational transform and far away from the axis, as in a), is a cubic a poor fit.}
    \label{fig:dBdl^2vsB_QUASR}
\end{figure*}

\begin{figure}
    \centering
    \includegraphics[width=0.5\textwidth]{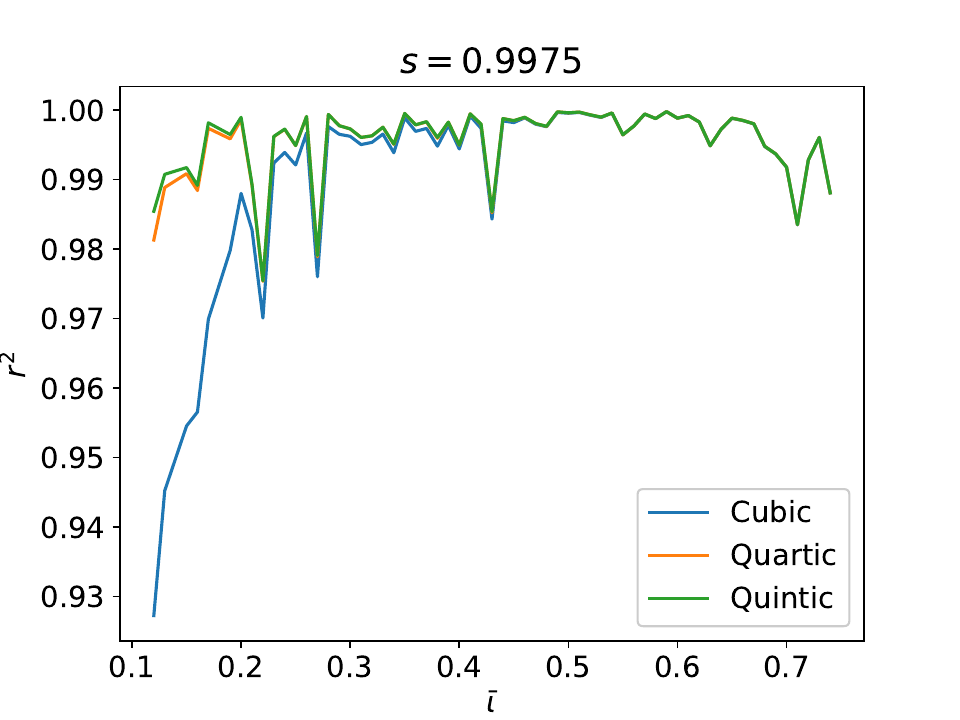}
    \caption{The coefficient of determination on the outermost surface, for different parameterizations of $(\partial_l B)^2$ as a function of $B$, across the Buller configurations, where each configuration has a different mean rotational transform. }
    \label{fig:coefficient of determination scan of iota}
\end{figure}

\begin{figure*}
    \centering
        \includegraphics[width=\textwidth]{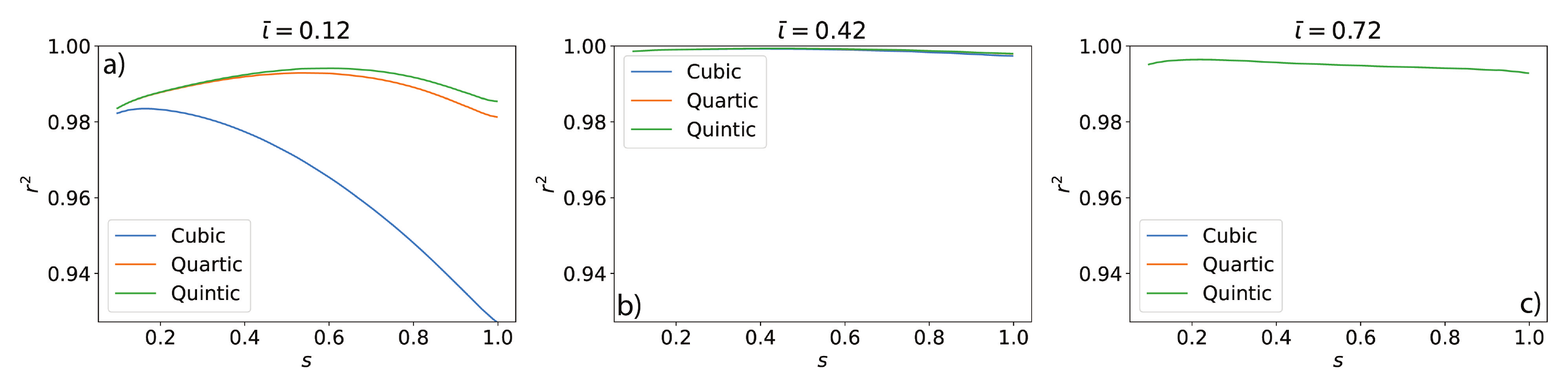}
    \caption{The coefficient of determination as a function of the surface, for three different mean rotational transforms, for different parameterizations of $(\partial_l B)^2$ as a function of $B$.}
    \label{fig:coefficient of determination scan of surf}
\end{figure*}

\begin{figure}
    \centering
    \includegraphics[width=0.5\textwidth]{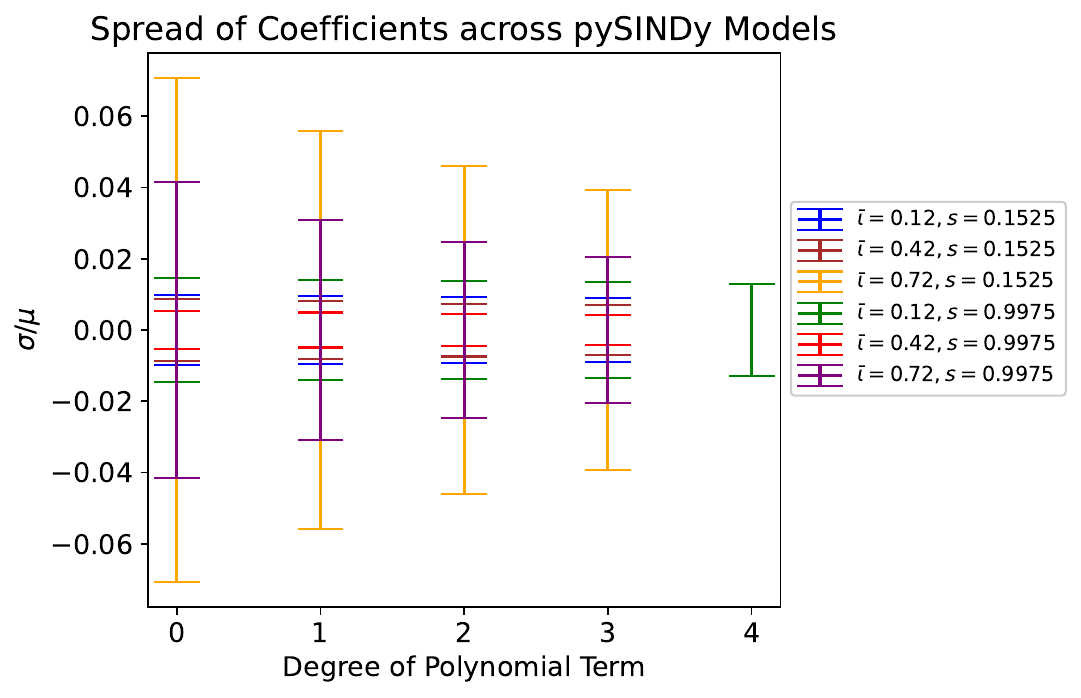}
    \caption{200 pySINDy models for each shown configuration and surface were produced, each which sampled, with replacement, $60\%$ of $1600$ available points on each flux surface. For each model, the standard deviation spread of the coefficients (normalized by the mean) is plotted.}
    \label{fig:pySINDy variance}
\end{figure}

\begin{figure*}
    \centering
    \includegraphics[width=\textwidth]{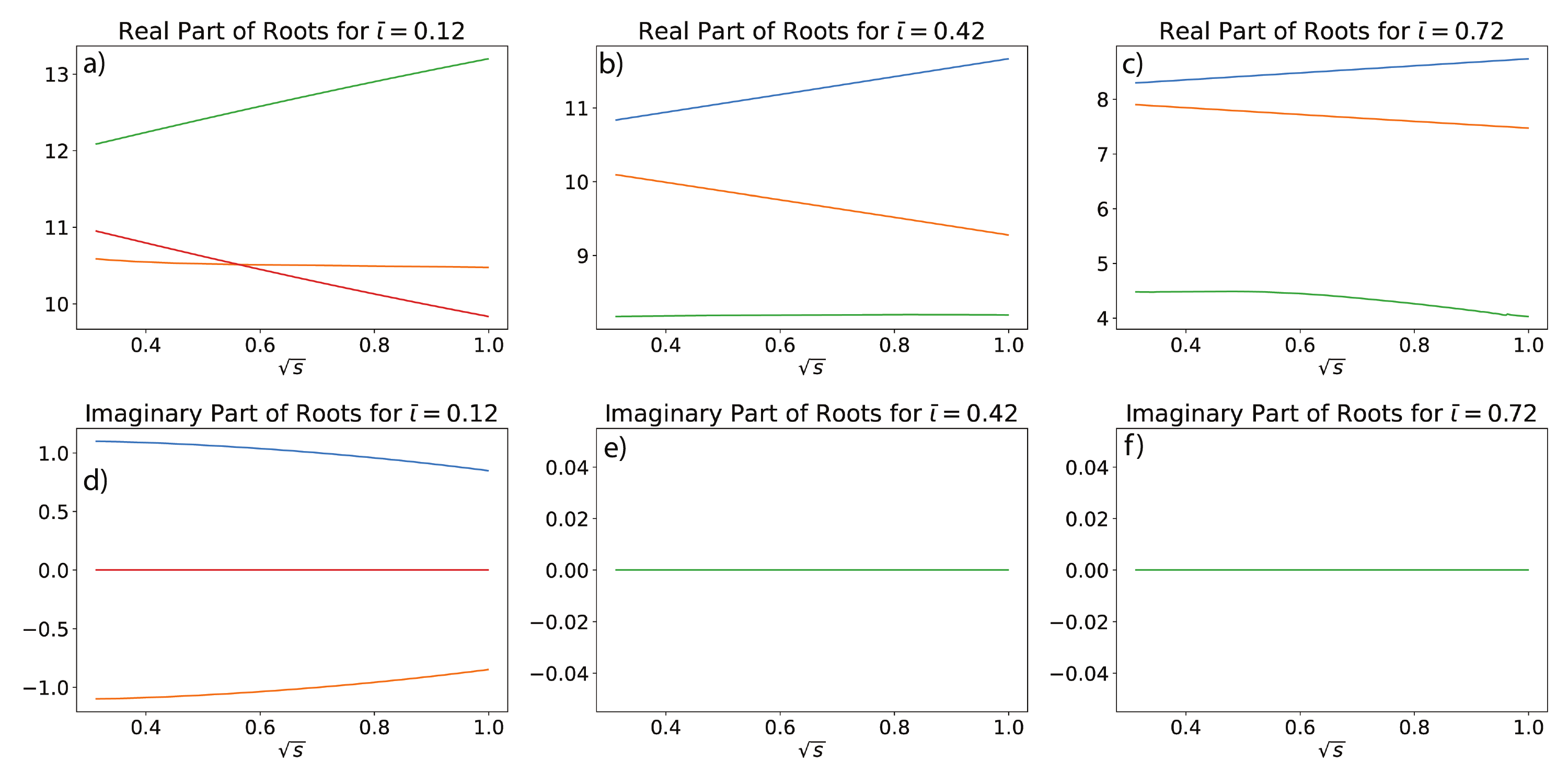}
    \caption{$(\partial_l B)^2$ is fit to a polynomial of $B$ across all flux surfaces -- a cubic in all cases except a), which is fit to a quartic as per Fig. \ref{fig:dBdl^2vsB}. Shown is the evolution of the roots.}
    \label{fig:roots}
\end{figure*}

We have put forth several analytical arguments in favor of the connection between QS and the well-known integrable PDEs, KdV, and Gardner's equation (for small rotational transforms). 
We have also demonstrated numerically good agreement with several configurations with good QS. In this Section, we take a different approach based on data-driven techniques to determine the underlying system of equations describing quasisymmetric $B$ on a flux surface. Our primary tool is pySINDy \cite{desilva2020, Kaptanoglu2022}, which fits governing differential equations and favors fewer terms to combat over-fitting. 

We have primarily used the family of quasi-axisymmetric VMEC equilibria generated in Ref.~\cite{buller2024family}. For additional verification, we considered further configurations from the QUASR database \cite{giuliani2023direct}. Henceforth, we refer to them as the Buller and the Giuliani configurations. For the latter, configurations were selected with three representative mean rotational transforms. They were chosen to have the number of field periods, $N_{fp}=2$, an aspect ratio close to six (to mirror the Buller configurations), and the lowest possible quasi-axisymmetry error (as defined in \cite{giuliani2023direct}).

Using the Buller configurations, we verified the quasisymmetry-imposed traveling wave form \eqref{eq:fund_B_alpha_B_ell_reln}. We calculated $H=-\partial_\alpha B / \partial_l B$, taking $l=0$ at $\phi=0$ and considering one field period. The traveling waveform could then be verified by the fact that $H$ was independent of $l$. Equivalently, we calculated $t=\int_0^\alpha d\alpha ' H$ and plotted $B$ on a surface parameterized by $l$ and $t$, in which case equation \eqref{eq:fund_B_alpha_B_ell_reln} indicates that we should see traveling waves of unit speed. These results are presented in Fig.~\ref{fig:traveling wave}, where we can clearly see QS manifest as a traveling wave. With the traveling waveform, the PDE governing $B$ reduces to ODEs, which pySINDy can easily determine. In fact, we also attempted to use pySINDy to identify PDEs directly, and could straightforwardly recover the traveling wave. Even in configurations with relatively substantial deviations from quasi-symmetry, the traveling wave structure is quite robust. Under the traveling wave assumption, the equations are reduced to ODEs anyway, so we focused on the ODE discovery problem. Future work can try to use pySINDy for PDE identification in regimes where significant deviations from traveling wave behavior are observed, which might require configurations quite far from quasi-symmetry.

Next, we use pySINDy to discover the underlying ODEs that govern $B$ on a flux surface. To that end, a sequentially thresholded least squares algorithm~\cite{brunton2016discovering,zhang2019convergence} was used for the fitting, with a cost function that promotes sparsity in the data fit. For each magnetic configuration's flux surface, the $B$ against $l$ profiles of multiple field lines spanning the surface were used as training data. The quantity $(\del_\ell B)^2$ on a flux surface was fitted against a linear combination of terms from a `feature library,' chosen here to be a selection of polynomial terms, with the model parameters as the coefficients. Since the KdV and Gardner's equation, together with the traveling waveform, lead to cubic and quartic polynomials, using the polynomial basis in pySINDy was a natural choice. A range of models was produced at varying sparsity parameters; the models were then tested on data from magnetic field lines not used in the training data. For sparsity parameters that were too large, it was found that there was a sharp drop in the success of the fit as too many terms were dropped; we chose the pySINDy model before this drop in performance. This is a version of hyperparameter scanning to find the Pareto-optimal fit (optimal tradeoff of model complexity versus model accuracy) to the data.

In Fig. \ref{fig:dBdl^2vsB}, we demonstrate the outcome of the pySINDy fit to $(\del_\ell B)^2$ on a flux surface for three representative configurations. A cubic was a good fit for all the configurations on flux surfaces sufficiently close to the axis, as expected from the NAE analysis. 


For configurations with low mean rotational transform, a cubic was no longer the best model on the outermost surfaces (see Figures \ref{fig:coefficient of determination scan of iota} and \ref{fig:coefficient of determination scan of surf}); pySINDy indicated that a quartic was an effective fit on such surfaces. We also see that a quintic performs well. However, within the error range, the benefit of using a quintic is very marginal compared to the quartic. 
We see in Fig. \ref{fig:dBdl^2vsB_QUASR} that the Giuliani configurations display similar behavior with deviation from a cubic fit away from the axis for configurations with low mean-$\iota$.
In Fig. \ref{fig:pySINDy variance}, we see that the deviation from the mean from many pySINDy models generated by sub-sampling is small. This provides additional evidence that cubic models fit the data extremely well, except near the plasma boundary in low-iota configurations, for which a quartic correction is robustly picked up during model identification.

Finally, the evolution of the roots in the representative sample configurations and surfaces is shown in Fig. \ref{fig:roots}. For the cubic case described by KdV, there are three real roots, as we discussed earlier in Section \ref{sec:numerical_verification}. The case for low $\iota$, described by Gardner's equation, on the other hand, can have two real and two complex conjugate roots owing to its quartic nature. In the leftmost panel of Fig. \ref{fig:roots}, we show the two real roots of the quartic $(\del_\ell B)^2$ given by the maximum and the minimum of $B$ on a flux surface $s$ and the other pair of complex conjugate roots.

\section{Discussions}
In summary, we have demonstrated that periodic soliton solutions of the KdV equation lead to a broad class of excellent quasisymmetric $B$. We show that under three very general assumptions, namely, periodicity in $\ell$, analyticity of $B$, and the existence of a nontrivial TW frame, we can motivate why QS naturally leads to the KdV or Gardner's equation for the magnetic field strength. The essential property is the Painlev\'e property, which ensures $B$ stays single-valued. Several results follow from the connection between QS and the KdV equation. 

Firstly, the description of a class of precisely quasisymmetric $B$ can be reduced to understanding the behavior of the roots $B_i$ of a cubic polynomial. These roots correspond to extrema or saddle points of $B$ on a surface. 
In particular, $B$, which satisfies the periodic KdV equation, is the potential of the periodic Schrodinger equation, where $B_i$'s are related to the endpoints of the band gap \citep{kamchatnov2000nonlinear,novikov1984theory_solitons}. 
Since the spectrum of KdV is time-independent, the roots do not depend on time.
This confirms the well-known statement that the extrema or saddle points of $B$ depend only on the flux label in QS \cite{Helander2014}. 
Therefore, the invariance of $B_X$ on a flux surface effectively implies that the rotational transform $\iota(\psi)$ must be independent of the field line label (time). In other words, QS leads to an integrable field line Hamiltonian, which preserves nested flux surfaces \citep{Helander2014}. Furthermore, since various aspects of the flux surface shaping can be deduced \citep{rodriguez2022_thesis} from the Fourier coefficients of $B$, the hidden lower dimensionality is beneficial for understanding shaping. A similar hidden lower dimensionality is found in the description of the curvature of isodynamic magnetic field lines \citep{schief2003nested}. We repeat for emphasis that these hidden lower dimensionalities in the field strength or geometry are a direct consequence of the underlying soliton dynamics: the KdV equation in the case of QS and the vortex filament equation in the case of isodynamic fields. 

Secondly, the hierarchy of quasisymmetric invariants, as given by \eqref{eq:integral_inv_QS_omni}, can be understood in terms of the hierarchy of the conserved quantities of the KdV equation. Evolution under the KdV equation guarantees the field-line independence of these quantities. 

Thirdly, we can estimate the maximum toroidal volume that QS allows solely from the properties of the roots and independent of any particular geometry. The overlap of the third root with the second could signal the breakdown of periodicity of $B$, as demonstrated numerically. At this radius, a separatrix could form, which could potentially be used as the basis for a non-resonant divertor \citep{boozer2015, punjabi2020}. 

Finally, the connection length can be much larger than the major radius, typically assumed in the literature. A longer connection length may have important implications for linear and nonlinear stability in devices with QS.

Our final approach is data-driven. Utilizing the vast amount of data on quasisymmetric equilibria generated in \cite{buller2024family} and the QUASR database \cite{giuliani2023direct}, and the state-of-the-art tool pySINDy, which uses sparse regression techniques to discover underlying models, we recover KdV and Gardner's equation. Thus, the Painlev\'e property that demands that $(\del_\ell B)^2$ can be at most a quartic in $B$ is verified accurately. This data-driven approach has the advantage that it can be applied to any configuration (including when quasi-symmetry is not relevant) to find low-dimensional ODE or PDE models underlying the data. 

Given the well-known robust stability of soliton solutions \cite{tao2009solitons_stable}, the spectral stability of the cnoidal waves \citep{bottman2009kdv}, and the robustness of the single-valuedness criterion, which led to the Painlev\'e property, this perhaps explains why any numerical optimizer trying to find good QS using the 2-term form or the triple-product form will most likely find only cubic or quartic terms in $(\del_\ell B)^2$.

\acknowledgments
We thank P. Helander, B. Khesin, J.W. Burby, M. Wheeler, J. Meiss, N. Duignan, V. Duarte, H. Weitzner, E. Rodriguez, F.P. Diaz and M. Landreman for helpful discussions and suggestions. This research was supported by a grant from the Simons Foundation/SFARI (560651, AB) and DoE Grant No. DE-AC02-09CH11466. 

\clearpage
\appendix

\section{The QS condition $\uD B=0$ in Clebsch variables}
\label{app:H}
In the Clebsch coordinates $(\psi,\alpha,\ell)$ with $\BD\ell =B$, the operators $\BD, \uD$ take the form
\begin{align}
    \BD = B \del_\ell, \quad \uD= \uD \alpha\; \del_\alpha  + \uD \ell\; \del_\ell
    \label{eq:uD_BD_Clebsch}
\end{align}
It follows from the expression for $\u$ given in \eqref{eq:QS_vector_u} that $\uD \alpha=-1$. Defining $H\equiv \uD \ell$, we can express the QS condition $\uD B=0$ as
\begin{align}
    \uD B = - \del_\alpha B + H \del_\ell B =0.
    \label{eq:QS_condition_uDB_Clebsch}
\end{align}
We can use the fact that the commutator $[\uD,\BD]=0$ and the QS condition \eqref{eq:QS_condition_uDB_Clebsch} to show that 
\begin{align}
    \del_\ell H=0, \quad \text{i.e.,}\quad H=H(\psi,\alpha).
\end{align}

In axisymmetry, the expressions for the vectors $\B,\u$ in standard cylindrical coordinates $(R,\phi,z)$ are given by
\begin{align}
    \B= F(\psi)\dl \phi + \dl \phi \times \dl \psi, \quad \u=R^2 \dl \phi,
    \label{eq:B_axis}
\end{align}
where $\iota, \psi$ denotes the rotational transform and the poloidal flux. It is convenient to define a flux coordinate  \citep{haeseleer_flux_coordinates} $(\psi,\theta,\phi)$ such that the Jacobian $1/\sqrt{g}=\dl \psi\times \dl \theta \cdot \dl \phi = F/R^2$. In the flux coordinates, we have
\begin{subequations}
\begin{align}
\B = \dl \psi \times \dl \alpha, \quad \alpha = \iota^{-1}\theta - \phi,\\
\BD = \frac{1}{\sqrt{g}}(\del_\phi + \iota \del_\theta ), \quad \u \cdot \dl = \del_\phi.
\end{align}
\end{subequations}

It is easy to check that $\BD,\uD$ commute since $\del_\phi \sqrt{g}=0$ in axisymmetry.

Now, we go to the Clebsch coordinates $(\psi,\alpha,\ell)$ to express $H$ in terms of the flux coordinate system we just introduced. The quantity $\ell$ can be determined by the following ODE describing a field line 
\begin{align}
    \frac{d\ell}{B}= \frac{d\phi}{\BD \phi}= \frac{d\theta}{\BD \theta}.
\end{align}
Utilizing $d\theta=\iota d\phi, \BD \phi = 1/\sqrt{g}$ and integrating along the field line keeping $\psi,\alpha$ fixed, we get
\begin{align}
    \ell = \frac{1}{\iota}\int_{\psi,\alpha} B \sqrt{g} d\theta,
\end{align}
up to an overall function $h(\psi,\alpha)$, which can be set to zero by appropriately choosing the lower bound of the integral. Changing the coordinates from $(\psi,\theta,\phi)$ to $(\psi,\alpha,\theta)$ such that
\begin{align}
 \phi=\iota^{-1}\theta-\alpha, \;\;
    \BD = (\iota/\sqrt{g}) \del_\theta , \;\; \uD = - \del_\alpha,
\end{align}
we can calculate the quantities $\BD \ell$ and $ \uD \ell$. It is easily checked that $\BD \ell=B$ as it should be. On the other hand, $\uD \ell$ and hence $H$ vanish, since a perfectly axisymmetric $B$ and $\sqrt{g}$ depend only on $\psi,\theta$. 

A similar analysis can be carried out in standard Boozer coordinates $(\psi,\vartheta_B, \phi_B)$ \cite{landreman2018a} where
\begin{subequations}
    \begin{align}
    \B=\dl\psi \times \dl \vartheta_B + \iota_N \dl \phi_B \times \dl \psi,\\
    \B = (G+N I)\dl \phi_B + I \dl \vartheta_B + K \dl \psi,\\
    1/\sqrt{g}=B^2/(G+\iota I), \quad \iota_N =\iota -N
\end{align}
\label{eq:Boozer_coords_B}
\end{subequations}
If $B$ is quasisymmetric, $B=B(\psi,\vartheta_B)$ in Boozer coordinates. Following the same steps as before, we obtain in $(\psi,\alpha,\vartheta_B)$ coordinates
\begin{subequations}
    \begin{align}
    \BD = \frac{\iota_N}{\sqrt{g}}\del_{\vartheta_B}, \; \uD = -\del_\alpha,\\
    \ell = \frac{1}{\iota_N}\int_\alpha \sqrt{g} B d\vartheta_B.
\end{align}
\end{subequations}
If QS is perfect, $\uD \ell$ would perfectly vanish because the Jacobian of Boozer coordinates is a function of $B$ and flux. However, for an approximately quasisymmetric system, the Jacobian depends on $\alpha$. Consequently, the $\uD \ell$ term does not vanish and can introduce a nonzero $\alpha$ dependent $H$. 

\section{Equivalence of \eqref{eq:fund_B_alpha_B_ell_reln} and the two-term form of QS}\label{sec:equivalence_2term}
We first point out the equivalence of \eqref{eq:fund_B_alpha_B_ell_reln} and the two-term relation. Since the QS condition has not appeared directly in the form \eqref{eq:fund_B_alpha_B_ell_reln} to the best of our knowledge, let us point out its relation to one of the standard forms of QS.
To impose QS on the vacuum field, we can use the two-term form \eqref{eq:2term_form}. We find that
\begin{align}
    \dl\psi \times \dl \lbr \Phi + \frac{F(\psi)}{G_0}\alpha\rbr\cdot \dl B =0
    \label{eq:2term_vacuum_B}
\end{align}
Thus, for vacuum fields, $B$ satisfies a traveling-wave-like solution \citep{sengupta_Paul2021vacuum} in $(\Phi,\alpha)$ of the form \cite{sengupta_Paul2021vacuum}
\begin{align}
    B=B(\Phi + \alpha F(\psi)/G_0,\psi)
    \label{eq:mod_B_TW_Phi_alpha}
\end{align}
Equivalently, $B=B(\vartheta,\psi)$ in Boozer coordinates.

On the other hand, eliminating $H$ from \eqref{eq:fund_B_alpha_B_ell_reln} using \eqref{eq:H_del_ell_del_alpha} we find that
\begin{align}
    \{\Phi + \alpha F(\psi)/G_0, B\}_{(\alpha,\ell)}=0 \Leftrightarrow \{\vartheta_B, B\}_{(\alpha,\ell)}=0
    \label{eq:PB_thetaB_alpha_ell}
\end{align}
where $\{ , \}$ denotes the usual Poisson bracket with respect to $(\alpha,\ell)$. Therefore,() is equivalent to the two-term form, while its $\ell$ derivative yields the relation \eqref{eq:fund_B_alpha_B_ell_reln}. Equation \eqref{eq:PB_thetaB_alpha_ell} also implies the equivalence of the TWs \eqref{eq:Boozer_coord_def} and \eqref{eq:dalpha_B_TW_B} in the Boozer and Clebsch coordinate systems.

%

\section{Additional configurations that satisfy the integrated KdV equation (12)}\label{app:additional_fig}
Here, we present figures similar to Figs. 1 and 3 in the main text, using data from different stellarator devices.

The precise QA configuration with larger negative shear in Fig.~\ref{fig:ansatz_validation_sQA} was obtained by a series of optimizations starting from the precise QA.\cite{landreman2021}
The mean shear $\bar{\hat{s}}$ was extracted from the precise QA configuration, and a new least-squares term $(\bar{\hat{s}} - \bar{\hat{s}}_*)^2$ was added to the objective function, where the target value for mean shear $\bar{\hat{s}}_*$ is taken to be slightly different from the extracted value. 
This new objective is then optimized locally, using the precise QA as the initial state. This ensures that the additional least-squares term is almost zero, so the precise QA is approximately an optimum to the modified objective function. Hence, the new local optimum can be expected to be close to the initial configuration. This process is then repeated starting from the slightly modified optimum, each step generating a similar equilibrium with a slightly different mean shear.

Mean shear is here defined as $b/\bar{\iota}$, where $b$ is the result of fitting a first-order polynomial $a+bs$ to the iota profile, where $s$ is the normalized toroidal flux; $\bar{\iota}$ is the rotational transform averaged over all radii. The mean shear in the original precise QA is about $-0.018$, and we rounded this to $-0.02$ incremented the target shear of each optimization in steps of $0.005$ to generate configurations with a range of shear. Fig.~\ref{fig:ansatz_validation_sQA} and Fig.~\ref{fig:ansatz_validation_psQA} show example with negative and positive shear, respectively. 

\begin{figure}[!h]
    \centering
    \includegraphics[width=0.45\textwidth]{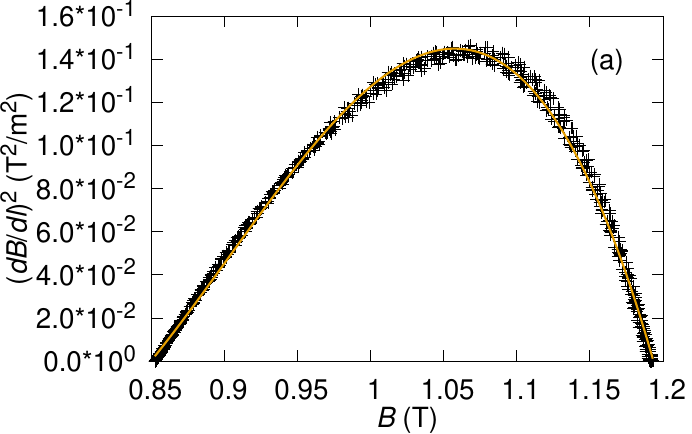 }
    \includegraphics[width=0.45\textwidth]{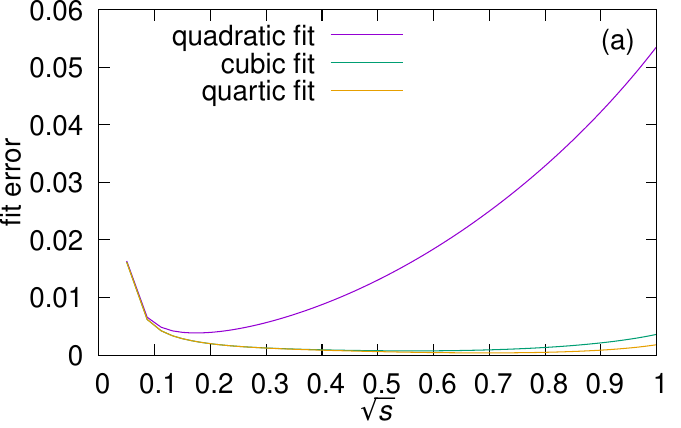 }
    \includegraphics[width=0.45\textwidth]{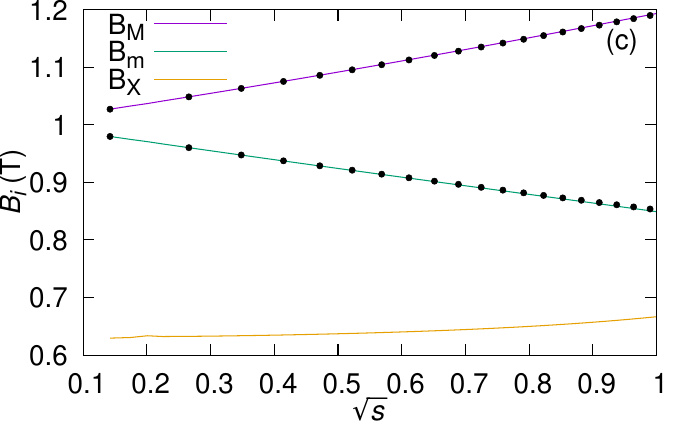 }
    \caption{The value of $(dB/d\ell)^2$ on the outermost flux surface in the precise QH \cite{landreman2021} as a function of $B$, alongside a cubic polynomial fit (a), and the errors of quadratic, cubic, and quartic fits on each flux surface (b). Panel (c) shows the behavior of the roots in the precise QH.}
    \label{fig:ansatz_validation_pQH}
\end{figure}

\newpage

\begin{figure}
    \centering
    \includegraphics[width=0.45\textwidth]{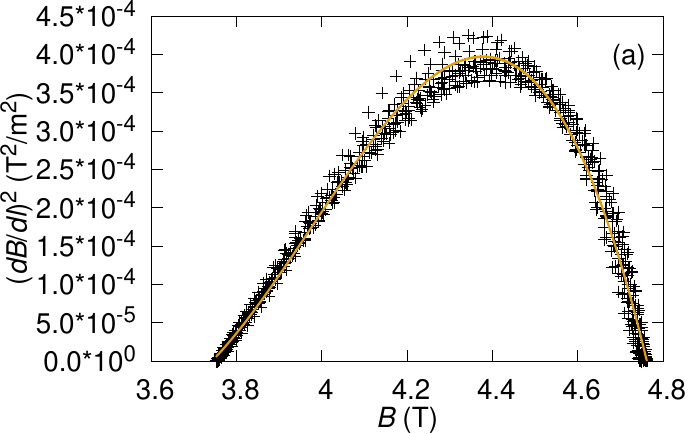 }
    \includegraphics[width=0.45\textwidth]{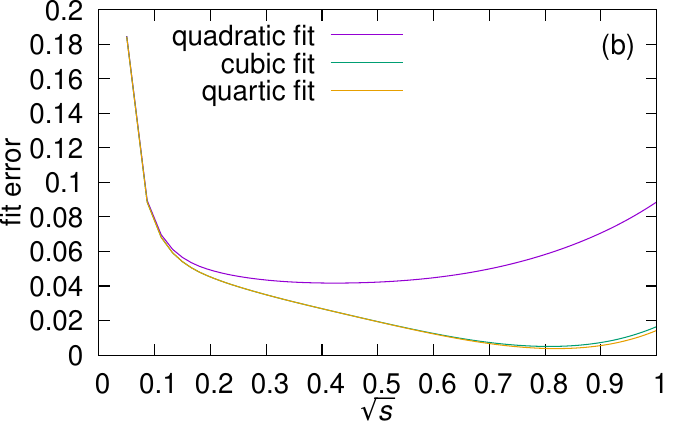 }
    \includegraphics[width=0.45\textwidth]{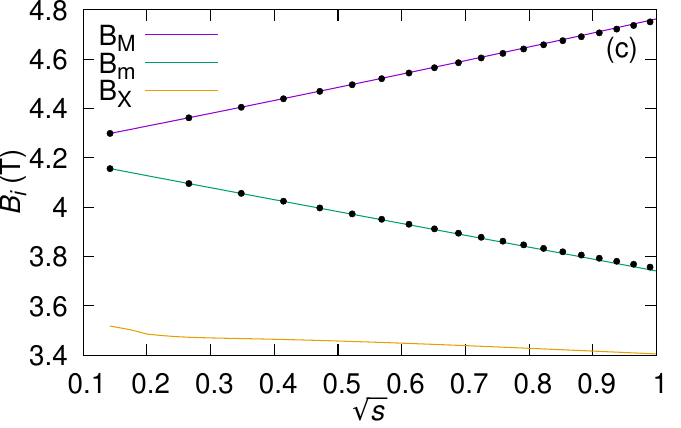 }
    \caption{The value of $(dB/d\ell)^2$ on the outermost flux surface in a precise QA version with average shear -0.08 as a function of $B$, alongside a cubic polynomial fit (a), and the errors of quadratic, cubic, and quartic fits on each flux surface (b). Panel (c) shows the behavior of the roots in this version of the precise QA.}
    \label{fig:ansatz_validation_sQA}
\end{figure}

\newpage

\begin{figure}[H]
    \centering
    \includegraphics[width=0.45\textwidth]{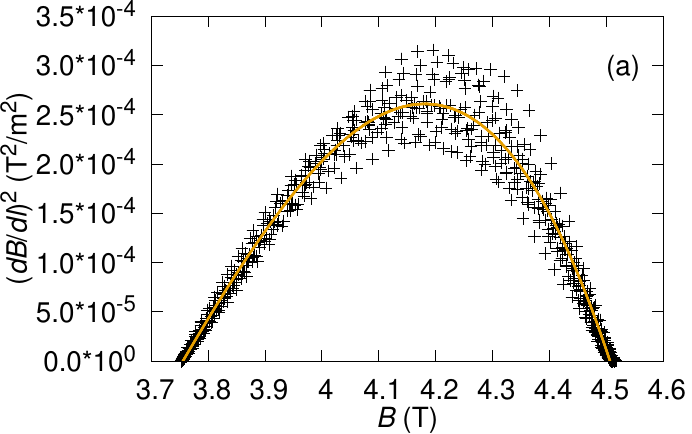 }
    \includegraphics[width=0.45\textwidth]{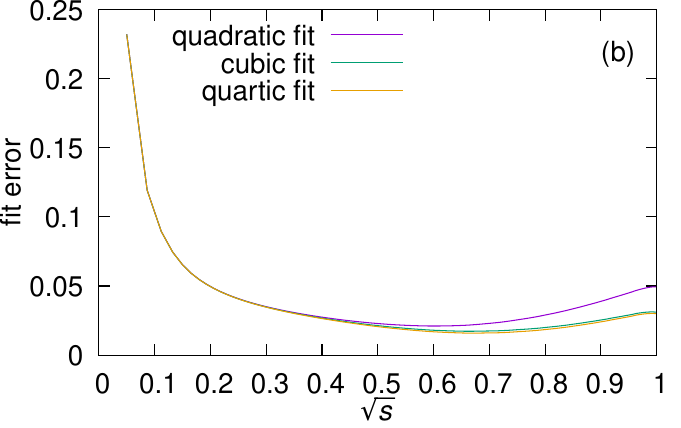 }
    \includegraphics[width=0.45\textwidth]{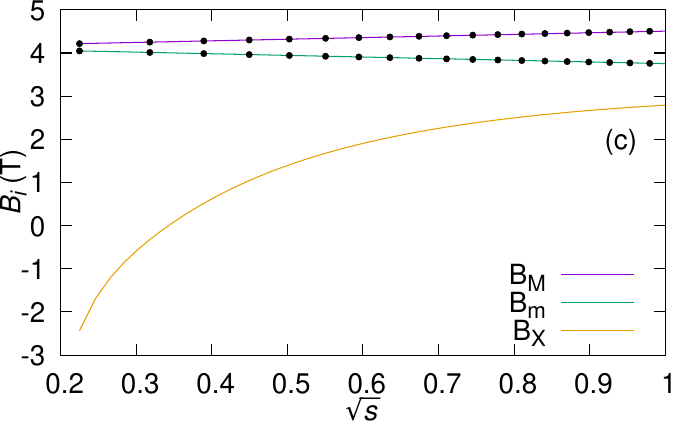 }
    \caption{The value of $(dB/d\ell)^2$ on the outermost flux surface in a precise QA version with average shear 0.04 as a function of $B$, alongside a cubic polynomial fit (a), and the errors of quadratic, cubic, and quartic fits on each flux surface (b). Panel (c) shows the behavior of the roots in this version of the precise QA.}
    \label{fig:ansatz_validation_psQA}
\end{figure}
\newpage
\vspace{-0.5cm}

\begin{figure}[H]
    \centering
    \includegraphics[width=0.45\textwidth]{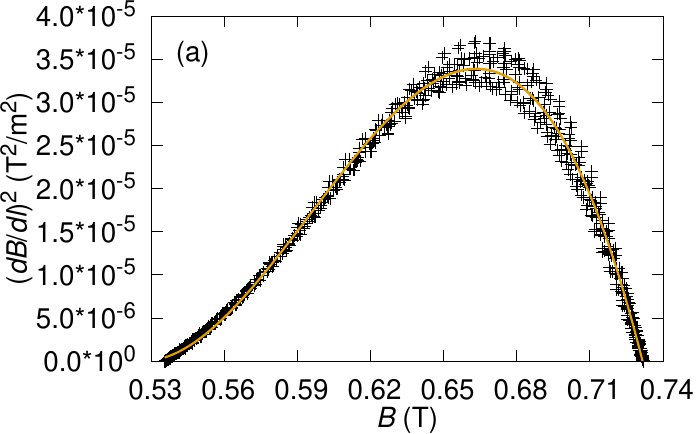 }
    \includegraphics[width=0.45\textwidth]{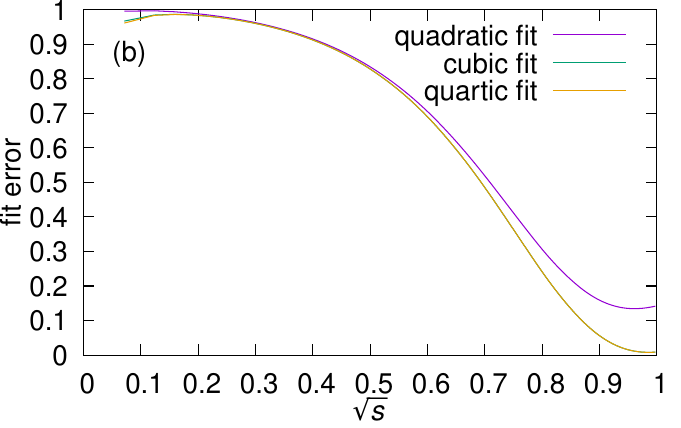 }
    \includegraphics[width=0.45\textwidth]{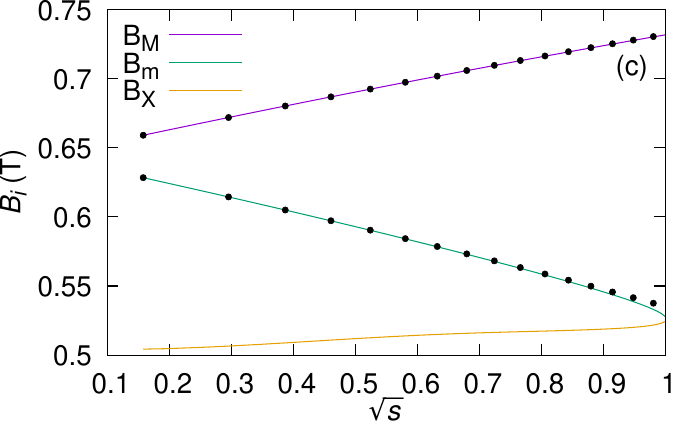 }
    \caption{The value of $(dB/d\ell)^2$ on the outermost flux surface in the QA equilibrium that was used in Fig. 2 in the main text as a function of $B$, alongside a cubic polynomial fit (a), and the errors of quadratic, cubic, and quartic fits on each flux surface (b). Panel (c) shows the behavior of the roots in this QA equilibrium.}
    \label{fig:ansatz_validation_from_f2}
\end{figure}

\begin{figure}[H]
    \centering
    \includegraphics[width=0.45\textwidth]{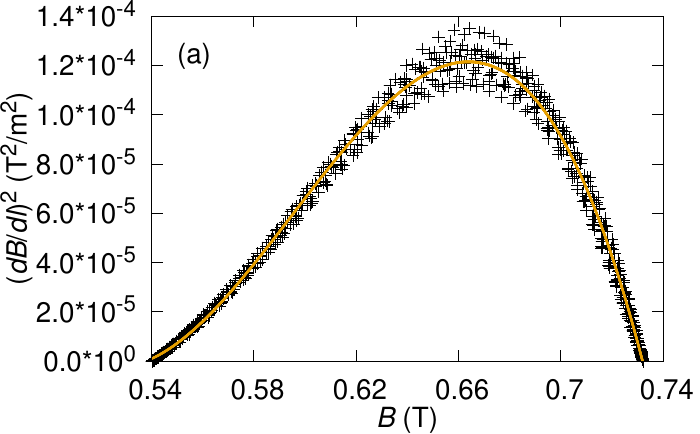 }
    \includegraphics[width=0.45\textwidth]{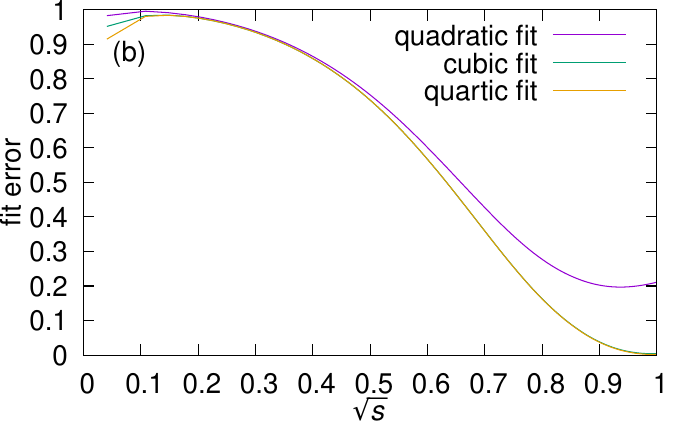 }
    \includegraphics[width=0.45\textwidth]{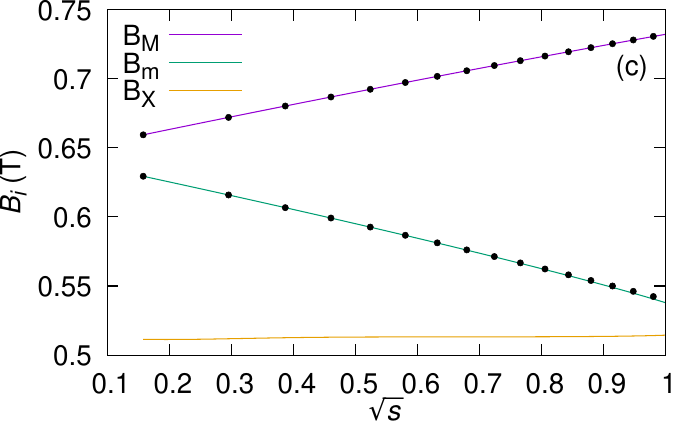 }
    \caption{The value of $(dB/d\ell)^2$ on the outermost flux surface in a QA equilibrium similar to that in Fig. \ref{fig:ansatz_validation_from_f2} as a function of $B$, alongside a cubic polynomial fit (a), and the errors of quadratic, cubic, and quartic fits on each flux surface (b). Panel (c) shows the behavior of the roots in this QA equilibrium.}
    \label{fig:ansatz_validation_i0p1}
\end{figure}

We show equilibria in Figures \ref{fig:ansatz_validation_from_f2} and \ref{fig:ansatz_validation_i0p1} that were not derived from the precise QA/QH. Instead, they were only optimized for quasisymmetry on the outermost flux surface, with the quasisymmetry degrading as the axis is approached. For both cases, the ratio of symmetry-breaking modes to quasisymmetric modes near the axis is $\sim 10^{-3}$. While this may seem a small value, the resulting ripple creates enough noise in the $(dB/d\ell)^2$ vs. $B$ plot to render it meaningless, and the error for all fits is $\sim 1$ in that region. The symmetry-breaking modes were filtered out when calculating the roots shown in panels (c). Note that the equilibrium shown in Fig. \ref{fig:ansatz_validation_from_f2} is the same one used in Fig. 2 in the main text.

\hrulefill
\bibliography{dBdl_PoP}

\end{document}